\documentclass[a4paper,nopacs,12pt]{iopart}
\usepackage{amssymb,epsfig,bm}
\usepackage{graphicx,iopams,setstack}
\usepackage{psfrag,textcomp}
\usepackage{upgreek} 
\newtheorem{prop}{Observation}

\newcommand{\bracket}[1]{\left\langle #1\right\rangle}

\newcommand{\beeq}[1] {\begin{eqnarray}#1\end{eqnarray}}
\newcommand{\beeqn}[1] {\begin{eqnarray*}#1\nonumber\end{eqnarray*}}

\begin{document}
\title[Spectral density of  products of Wishart dilute random matrices. Part I]{Spectral density of  products of Wishart dilute random matrices. Part I: the dense case}
\author{Thomas Dupic}
\address{Department of Mathematics, King's College London, Strand, London WC2R 2LS, United Kingdom}
\author{Isaac P\'erez Castillo}
\address{Department of Mathematics, King's College London, Strand, London WC2R 2LS, United Kingdom}
\address{Departamento de Sistemas Complejos, Instituto de F\'isica, UNAM, P.O. Box 20-364, 01000 M\'exico D.F., M\'exico}
\begin{abstract}
In this work we study the spectral density of products of Wishart diluted random matrices of the form $X(1)\cdots X(M)(X(1)\cdots X(M))^T$ using the Edwards-Jones trick to map this problem into a system of interacting particles with random couplings on a multipartite graph. We apply the cavity method to obtain recursive relations in typical instances from which to obtain the spectral density. As this problem is fairly rich, we start by reporting in part I a lengthy analysis for the case of dense matrices. Here we  derive that the spectral density is a solution of a polynomial equation of degree $M+1$ and obtain exact expressions of it for $M=1$, $2$ and $3$.  For general $M$, we are able to find the exact expression of the spectral density only when all the matrices $X(t)$ for $t=1,\ldots, M$ are square. We also make some observations for general $M$, based admittedly on some weak numerical evidence, which we expect to be correct.
\end{abstract}

\maketitle

\section{Introduction}
We can savely say that research in random matrix theory is alive and well. This is not unsurprising considering the ever expanding range of applications from Mathematics, to Physics, to  Economics, to Biology, to Engineering, and so on. The field is expanding so fast that it is becoming more and more difficult to be up to date or even trying to master all mathematical approaches available (SuSy, determinantal, spin glass techniques, free probability theory, etc).\\

\noindent From the various active research topics, there is a visibly recent and growing interest in understanding the statistical properties of products of random matrices \cite{Burda2010,Burda2010b,Akemann2013,Akemann2013b,Akemann2013c,Gudowska2003,Gotze2010} even though this topic had already been moderately active in the past \cite{Crisanti1993}. While a sizeable part of these studies deal with complex matrices, the purpose of this series of works (parts I and II) is to complement the already existing knowledge by extending the current studies to the realm of real and diluted random matrices \footnote{A good summary of these results was already presented by T. Dupic in a poster at the \textit{Mini-Conference on Statistical Mechanics of Glassy and Disordered Systems}, King's College London, 20-21 May 2013.}. As it has been noticed  in several occassions, such a simple imposition in the matrix entries (i.e. many zero entries proportional to the matrix size and few non-zero entries independent of the matrix size) complicates the mathematical analysis. In fact, the mathematical derivations have become so voluminous in the preparation of this work that we have preferred to split it into two parts: The first part corresponds to the analysis of the ensemble of matrices in the dense limit, while the second part corresponds to the analysis of general diluted matrices,  while paying particular attention on generalising the results of part I to the case of random regular graphs. We have decided to start with dense matrices for two reasons: firstly, it seems relatively ``the easiest case''; secondly, there are already quite some known results in this case -derived by different methods-. We expect that part I will be a nice introduction to the spin-glass methods we use and, in turn, will complement the existing alternatives derivations (see for instance \cite{Akemann2013,Burda2010,Burda2010b} and references therein), hopefully bringing new results as well.\\

\noindent This paper is organized as follows: in section \ref{sect:spp} we introduce the problem mathematically and see how the spectral density can be derived from the statistical properties encoded in a partition function. In section \ref{cmdl} we tackle the effective Hamiltonian by using the cavity method and perform the dense limit. We also show that the spectral density is given by one of the solutions of a polynomial equation of degree $M+1$. In section \ref{m123} we consider the cases for $M=1$, $2$, and $3$, which corresponds to those instances in which the polynomial equation can be solved by radicals, and compare our exact results with numerical diagonalisation. Section \ref{generalm} is a mixed bag of exact results and numerical obversations that lead to, what we believe are, exact results. The last section \ref{confw} is for conclusions and advancing what is to be expected in part II of this technical report.
\section{Spectral density as a spin glass problem}
\label{sect:spp}
While there exist various mathematical approaches to study statistical properties of matrix ensembles, as soon as these matrices are diluted (i.e. many zero entries and a few non-zero entries) our  choice of available techniques rapidly narrows. Unfortunately, this is the situation we encounter here, so we take the safe route of undertanding the spectral density as a spin glass problem and applying the well-understood techniques of the latter research area to say something about the original problem.\\
Let us therefore consider the following ensemble of matrices: given $M$ matrices $X(t)$ for  $t=1,\ldots, M$, we define the matrix product:
\beeq{
Y=X(1)\cdots X(M)(X(1)\cdots X(M))^T\,.
\label{my}
}
Here each matrix $X(t)$ is a $N(t)\times N(t+1)$ rectangular matrix. This implies that $Y$ is a positive-definite symmetric matrix of order $N(1)\times N(1)$. Denoting as $\{\lambda_i^Y\}_{i=1,\ldots,N(1)}$ its collection of real and positive eigenvalues, its spectral density is defined by
\beeq{
\rho_{Y}(\lambda)=\frac{1}{N(1)}\sum_{i=1}^{N(1)}\delta(\lambda-\lambda_i^{Y})\,.
\label{eq:density}
}
The next step is to relate the density \eref{eq:density} to some mathematical object one is familiar with in statistical mechanics of disordered systems. This is done by means of the Edwards-Jones trick \cite{Edwards1976}, which we briefly explain here. One first recall the following identity from distribution theory (or theory of generalised functions)
\beeqn{
\lim_{\epsilon\to 0^{+}}\frac{1}{x-i\epsilon}={\rm P}\frac{1}{x}+i\pi\delta (x)\,,
}
where $\delta(x)$ is the Dirac delta. The expression above must be understood as distributions (that is, in the strictest mathematical way, under the integration sign and being multiplied by a test function). Using also that $1/x=d \ln(x)/dx$ we can rewrite \eref{eq:density} as follows
\beeqn{
\rho_{Y}(\lambda)&=\lim_{\epsilon\to0^+}\frac{1}{\pi N(1)}{\rm Im}\frac{\partial}{\partial z}\sum_{i=1}^{N(1)} \ln(z-\lambda_i^{Y})\Big|_{z=\lambda-i\epsilon}\\
&=\lim_{\epsilon\to0^+}\frac{1}{\pi N(1)}{\rm Im}\frac{\partial}{\partial z} \ln \prod_{i=1}^{N(1)}(z-\lambda_i^{Y})\Big|_{z=\lambda-i\epsilon}\\
&=-\lim_{\epsilon\to0^+}\frac{2}{\pi N(1)}{\rm Im}\frac{\partial}{\partial z} \ln \left(\frac{1}{\sqrt{\det (z\bm{1}_{N(1)}- Y)}}\right)_{z=\lambda-i\epsilon}\,
}
where $\bm{1}_{N(1)}$ is the $N(1)\times N(1)$ identity matrix. In this derivation we have used the algebraic result $\prod_{i=1}^{n}\lambda_{i}^B=\det (B)$, with $\{\lambda_{i}^B\}_{i=1,\ldots,n}$ being the eigenvalues of a $n\times n$ matrix $B$. Notice also that we have rewritten $\ln \det (z\bm{1}_{N(1)}- Y)=-2\ln \left(1/\sqrt{\det (z\bm{1}_{N(1)}- Y)}\right)$.  Since we know that for a positive definite matrix $B$, the expression $1/\sqrt{\det (B)}$ can be expressed as a multi-dimensional Gaussian integral\footnote{It is also possible to stick with the expression $\ln \det (z\bm{1}_{N(1)}- Y)$ and use Grassmann variables to write down an integral expression for $\det (z\bm{1}_{N(1)}- Y)$.}, we end up having the following expression for the density in \eref{eq:density}:
\beeq{
\rho_{Y}(\lambda)&=-\lim_{\epsilon\to0^+}\frac{2}{\pi N(1)}{\rm Im}\frac{\partial}{\partial z} \ln  Z(z)\Big|_{z=\lambda-i\epsilon}\label{newd}\,,\\
Z(z)&=\frac{1}{\sqrt{\det (z\bm{1}_{N(1)}- Y)}}\nonumber\\
&=\int \left[\prod_{i=1}^{N(1)}\frac{dw_i(1)}{\sqrt{2\pi}}\right]\exp\left[-\frac{1}{2}\sum_{i,j=1}^{N(1)}w_i(1)(z\bm{1}_{N(1)}- Y)_{ij}w_j(1)\right]\,.\label{pf}
}
Expressions \eref{newd} and  \eref{pf} entice us to understand $\rho_Y(\lambda)$ as an observable corresponding to a system of $N(1)$ continuous spins $w_i(1)$ for $i=1,\ldots, N(1)$ interacting with the Hamiltonian
\beeq{
H[\bm{w}(1)]=\frac{1}{2}\sum_{i,j=1}^{N(1)}w_i(1)(z\bm{1}_{N(1)}- Y)_{ij}w_j(1)\,,
}
having a quenched random interaction matrix $z\bm{1}_{N(1)}- Y$ and unit temperature. Of course one should not forget that $z$ is generally complex, which strictly speaking implies that we do not have a Boltzmann measure. Yet, from a mathematical point of view, most (if not all) of the mathematical tricks that one can use in an interacting system defined by expressions \eref{newd} and  \eref{pf} still apply, so one refers formally to $Z(z)$ as the partition function (and the use of the jargon of \textit{thermal averaging}, \textit{Boltzmann measure}, et cetera swiftly follows).\\
Notice finally, that within this new context of interacting particles, the spectral density $\rho_Y(\lambda)$ is related to the thermal average of the second moment of variable $w_i(1)$. Indeed, from results \eref{newd} and  \eref{pf} we see that
\beeq{
\hspace{-2cm}\rho_Y(\lambda)&=\lim_{\epsilon\to 0^{+}}\frac{1}{\pi N(1)}{\rm Im}\sum_{i=1}^{N(1)} \bracket{w_i^2(1)}_{z=\lambda-i\epsilon}\,,\\
\hspace{-2cm}\bracket{\cdots}_z&=\frac{1}{Z(z)}\int \left[\prod_{i=1}^{N(1)}\frac{dw_i(1)}{\sqrt{2\pi}}\right] (\cdots)\exp\left[-\frac{1}{2}\sum_{i,j=1}^{N(1)}w_i(1)(z\bm{1}_{N(1)}- Y)_{ij}w_j(1)\right]\,.
}
So far this approach is valid for any symmetric matrix $Y$.\\
 We now turn our attention for $Y$ matrices with the structure as given in \eref{my}. Here it is important to notice that for matrices of this sort, its entries are correlated. It is possible to treat the problem as it is, but it is much easier to disentangle the correlation of the entries of $Y$ and express the problem in terms of the set of matrices $X(t)$. This is done by augmenting the system by adding new particles. The end result is the new thermal average \footnote{Here, we ignore the $2\pi$ factors as they are not important in the evaluation of the spectral density.}:
\beeq{
\hspace{-1cm}\bracket{\cdots}_z&=\frac{1}{Z(z)}\int \left[\prod_{t=1}^{M+1} d\bm{w}(t)\right](\cdots)e^{-H[\bm{w}(1),\bm{w}(M+1)]}\prod_{t=1}^{M}W_t[\bm{w}(t+1)|\bm{w}(t)]\,,\\
\hspace{-1cm}Z(z)&=\int \left[\prod_{t=1}^{M+1} d\bm{w}(t)\right](\cdots)e^{-H[\bm{w}(1),\bm{w}(M+1)]}\prod_{t=1}^{M}W_t[\bm{w}(t+1)|\bm{w}(t)]\,,
\label{nm}
}
where we have defined
\beeq{
W_t[\bm{w}(t+1)|\bm{w}(t)]&=\delta\left[\bm{w}(t+1)-X^{T}(t)\bm{w}(t)\right]\,,\\
H[\bm{w}(1),\bm{w}(M+1)]&=\frac{z}{2}\bm{w}^T(1)\bm{w}(1)-\frac{1}{2}\bm{w}^{T}(M+1)\bm{w}(M+1)\,,
}
with $\bm{w}(t)=(w_{1}(t),\ldots w_{N(t)}(t))$, and where $\delta(\bm{x})=\prod_{i=1}^n\delta (x_i)$ for $\bm{x}=(x_1,\ldots,x_n)$.\\
The new partition function \eref{nm} (and corresponding Boltzmann measure) is completely general for any set of rectangular matrices $X(t)$. However, due precisely to their rectangular caracter, finding a way to capture mathematically the most general ensemble embracing all possible scenarios seems a doomed task (e.g. how to properly define correlations between two different recangular matrices $X(t)$ and $X(t')$, or how to define symmetric properties of a given rectangular matrix $X(t)$). Thus, we consider the simpler ensemble characterised by a lack of correlation between any distinct pair of matrices $X(t)$ and $X(t')$. Besides, for each matrix $X(t)$ its non-zero entries $x_{ij}(t)$ are independent and identically distributed random variables with zero mean and variance $J^2(t)$. It is worth to point out that simple cases with correlation between matrices have been already studied \cite{Vinayak2013}. The extension of the present work to consider a non-zero correlation matrix $\Xi(t,t')=\bracket{X(t)X(t')}$ is currently under way \cite{Perez2014}.

\section{Cavity method and dense limit}
\label{cmdl}
In part I of this report we are going to use the cavity method\footnote{See, for instance, \cite{Mezard2009} for generalities of the cavity method or \cite{Rogers2008,Rogers2009} for simpler applications of the cavity method in similar problems. Another useful reference to understand the cavity method is \cite{Shamir2000}} to tackle the problem. For those interested in its replica counterpart, it will be used for completeness in part II. Note also that we are interested here in the properties of  such matrices in the thermodynamic limit, that is, when the size of the matrices tends to infinity. Finite size corrections to the spectral density are possible by using e.g. the methodology of \cite{Metz2014}, but we postpone this analysis for another occasion.\\

\begin{figure}[h]
\begin{center}
\begin{picture}(175,175)
\put(0,-5){$N(1)$}
\put(65,-5){$N(2)$}
\put(140,-5){$N(3)$}
\put(210,-5){$N(4)$}
\put(75,70){$i$}
\put(145,90){$j$}
\put(105,83){$x_{ij}(2)$}
\put(0,0){\includegraphics[width=8cm, height=6cm]{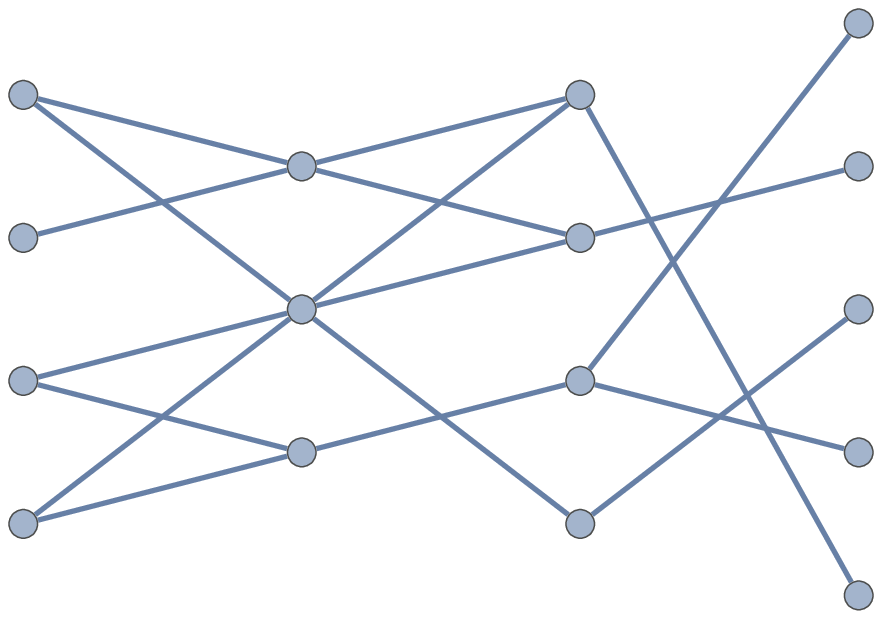}}
\end{picture}
\caption{Pictorial representation of a multipartite graph with $N(1)=4$, $N(2)=3$, $N(3)=4$ and $N(4)=5$}
\label{fig:matrix}
\end{center}
\end{figure}

\begin{figure}[h]
\begin{center}
\begin{picture}(175,175)
\put(5,20){$w_\ell$}
\put(-20,80){$q_{t+1}$}
\put(240,80){$k_{t+1}$}
\put(65,50){$x_\ell(t)$}
\put(105,100){$t+1$}
\put(107,70){$w$}
\put(0,0){\includegraphics[width=8cm, height=6cm]{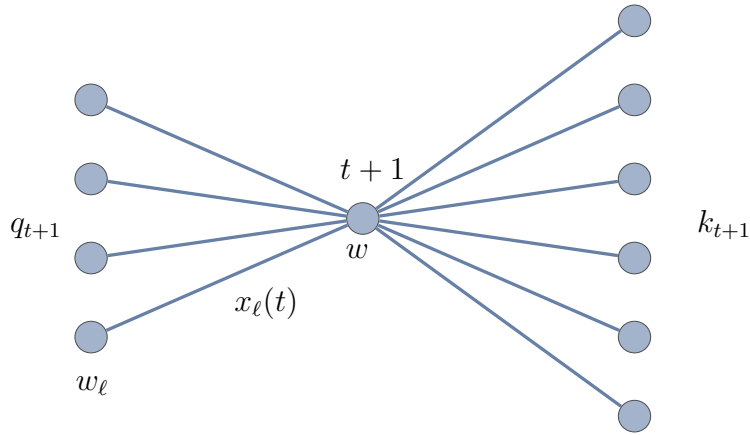}}
\end{picture}
\caption{Pictorial representation of a local node with variable $w$ at the $(t+1)$-layer in a multipartite graph}
\label{fig:local}
\end{center}
\end{figure}

Let us start be fixing some of the notation and giving some pictorial representation of the problem. We  first notice that each matrix $X(t)$ can be understood as a weighted bipartite graph in which a weighted link $x_{ij}(t)$ connects a node $i$ with $i=1,\ldots,N(t)$ with a node  $j$ with $j=1,\ldots, N(t+1)$.  For a node on the $t$-layer (or a $t$-node) $x_{ij}(t)$ can be understood as an outgoing link while  this will be considered as an incoming link for a $(t+1)$-node. As we have $M$ matrices $X(t)$, the overall graph is a combination of coupled bipartite graphs between consecutive $t$-layers (also called a multipartite graph), as depicted in figure \ref{fig:matrix}. 
Now, for each matrix $X(t)$ the total number of outgoing links in the $t$-layer must match the number of incoming links in the $(t+1)$-layer. Let us denote as $k_{i,t}$ the number of outoing links from node $i$ in the $t$-layer, and as $q_{j,t+1}$ the number of incoming links to node $j$ in the $(t+1)$-layer. We must have that $\sum_{i=1}^{N(t)} k_{i,t}=\sum_{j=1}^{N(t+1)} q_{i,t+1}$. Alternatively we write $N(t)\overline{k}_t=N(t+1)\overline{q}_{t+1}$, with $\overline{k}_t=(1/N(t))\sum_{i=1}^{N(t)}k_{i,t}$ and $\overline{q}_{t+1}=(1/N(t+t))\sum_{i=1}^{N(t+1)}q_{i,t+1}$ the average outgoing in incoming connectivities, respectively.\\
Consider next how the situation looks like around one of the $(t+1)$-nodes (see figure \ref{fig:local}). For simplicity we drop the indices $i,j$ and work with generic nodes. Here a generic node in the $(t+1)$-layer with variable $w$ has $q_{t+1}$ incoming links with weights $\{x_{\ell}(t)\}_{\ell=1}^{q_{t+1}}$ and $k_{t+1}$ outgoing links.
To apply the cavity method we focus in the generic case depicted in figure \ref{fig:local} and try to derive marginal distributions evaluating the constraint $\delta\left(w-\sum_{\ell=1}^{q_{t+1}} x_\ell(t) w_\ell\right)$. This constraint states that $q_{t+1}$ variables $\{w_\ell\}_{\ell=1}^{q_{t+1}}$ at the $t$-layer will determine the value of $w$ of the $(t+1)$-node. This value will in turn participate in $k_{t+1}$ other constraints.\\
 Suppose we fix the value of $w$ at the $(t+1)$-node and wonder about its impact to the left and to the right. Let us denote as $P_{t}$ the marginals at the $t$-layer resulting from this operation, while the marginals to the right will be denoted as $Q_{t+1}$. These marginals are obviously different as the role that the variabble $w$ at the $(t+1)$-node plays to the left and to the right is different. It is possible to write down recursive relations for them. Indeed, for the $Q$-marginals we have that
\beeq{
Q_{t}(w)&=\frac{1}{Q_t}\left[\int dw'\prod_{\ell=1}^{k_{t+1}} Q_{t+1,\ell}(w')\right]\left[\int\prod_{\ell=1}^{q_{t+1}-1}d w_\ell P_{t,\ell}(w_\ell)\right]\nonumber\\
&\times\delta\left(w'-x(t)w-\sum_{\ell=1}^{q_{t+1}-1} x_\ell(t) w_\ell\right)\,,
\label{eq:cav1}
}
for $t=1,\ldots, M-1$. Similarly, for the $P$-marginals one finds
\beeq{
\hspace{-2cm}P_{t+1}(w)=\frac{1}{P_{t+1}}\left[\prod_{\ell=1}^{k_{t+1}-1} Q_{t+1,\ell}(w)\right]\left[\int \prod_{\ell=1}^{q_{t+1}}d w_\ell P_{t,\ell}(w_\ell)\right]\delta\left(w-\sum_{\ell=1}^{q_{t+1}} x_\ell(t) w_\ell\right)\,,
\label{eq:cav2}
}
for $ t=1,\ldots, M-1$. These equations do not apply to the first and last layers as their neighbourhood is rather different: nodes in the first layer only have outgoing links, while nodes in the last layer only have incoming links. Moreover, the nodes in the last layer carry a weight $e^{w^2/2}$, as indicated by the partition function. Thus in this case the corresponding equations are:
\beeq{
P_{1}(w)&=\frac{e^{-\frac{z}{2}w^2}}{P_{1}}\prod_{\ell=1}^{k_1-1} Q_{1,\ell}(w)\nonumber\\
Q_M(w)&=\frac{1}{Q_M}\left[\int dw'e^{\frac{1}{2}(w')^2}\right]\left[\int\prod_{\ell=1}^{q_{M+1}-1}d w_\ell P_{M,\ell}(w_\ell)\right]\\
&\times\delta\left(w'-x(M+1)w-\sum_{\ell=1}^{q_{M+1}-1} x_\ell(M+1) w_\ell\right)\nonumber
\label{eq:begend}
}
The set of equations \eref{eq:cav1},\eref{eq:cav2}, and \eref{eq:begend} are our cavity equations for the $Q$ and $P$ marginals. Once they are solved, any statistical quantity of interest can be estimated from them. In particiular, the spectral density is related to the second moment of the actual marginals in the first layer. These are given by
\beeqn{
\mathcal{P}(w)&=\frac{e^{-\frac{z}{2}w^2}}{\mathcal{P}}\prod_{\ell=1}^{k_{1}} Q_{1,\ell}(w)
}
As it has been noticed in previous works \cite{Rogers2008,Rogers2009,Rogers2010}, the subset of Gaussian distributions is a functional fixed-point of the cavity equations. Thus, parametrising the cavity marginals as $Q_{t}(w)=\frac{1}{\sqrt{2\pi/(x^2(t)\gamma(t))}}e^{-\frac{x^2(t)\gamma(t)}{2}w^2}$ and $P_{t}(w)=\frac{1}{\sqrt{2\pi \sigma(t)}}e^{-\frac{w^2}{2\sigma(t)}}$ permits us to transform the original functional equations into standard equations for variances;
\beeq{
\hspace{-2.5cm}\gamma(t)&=\left( \sum_{\ell=1}^{q_{t+1}-1}x^2_\ell(t)\sigma_{\ell}(t)+\frac{1}{\sum_{\ell=1}^{k_{t+1}}x^2_\ell(t+1)\gamma_{\ell}(t+1)}\right)^{-1}\,,\quad t=1,\ldots,M-1\nonumber\\
\hspace{-2.5cm} \gamma(M)&=-\frac{1}{1- \sum_{\ell=1}^{q_{M+1}-1}x^2_\ell(M)\sigma_\ell(M)}\,,\nonumber\\
\hspace{-2.5cm}\sigma(1)&=\frac{1}{z+\sum_{\ell=1}^{k_1-1}x^2_\ell(1)\gamma_\ell(1)}\,,\nonumber\\
\hspace{-2.5cm} \sigma(t+1)&=\frac{1}{\sum_{\ell=1}^{k_{t+1}-1}x^2_\ell(t+1)\gamma_\ell(t+1)+\frac{1}{\sum_{\ell=1}^{q_{t+1}} x^2_\ell(t)\sigma_\ell(t)}}\,,\quad t=1,\ldots,M-1\,,
\label{eq:cava}
}
while for the actual variance, the one parametrising $\mathcal{P}$, we obtain $\Delta=[z+\sum_{\ell=1}^{k_1}x_\ell^2(1)\gamma_\ell(1)]^{-1}$. We keep the set of equations \eref{eq:cava} for a generic node as they are, even though they can easily be rewritten for a given instance of a multipartite graph.  These equations are of course valid for typical and large multipartite graph where short loops are rare, and can be solved numerically by using fixed point iteration methods (also called belief-propagation in this context) to render the spectral density of any large instance. Alternatively, one may prefer to use the ensemble equations and used population dynamics instead. This, and exact solutions for random regular multipartite graphs, will be the topic of part II of this work.\\
Here we focus on the dense case that, as we will see, it is already quite challenging to deal with. To perform the dense limit  $k_t,q_t\to\infty$ in \eref{eq:cava}, we first rescale the entries $x_\ell(t)\to x_\ell(t)/\sqrt{q_{t+1}}$. In the subsequent derivations we encounter expresions of the sort $\frac{1}{q_{t+1}}\sum_{\ell=1}^{q_{t+1}-1}x^2_\ell(t)\sigma_{\ell}(t)\to  J^2(t)\sigma(t)$. Here several steps have been done at the same time in the limit for large $q_{t+1}$ and $k_t$, the important ones being: (i) the variables $x^2_\ell(t)$ and $\sigma_\ell(t)$ decorrelate; (ii) no Onsager's reaction term is generated; (iii) nodes are equivalent within each layer $t$ \footnote{Onsager's reaction term is the correcting term that appears when determining the value of a random variable $x_i$, whose value depends e.g. on a sum of random variables in which the variable $x_i$ is also involved. If the typical value of the sum is zero, any change in $x_i$ will determine the sign of the sum and therefore one must account for changes in $x_i$ (thus Onsager's reaction term). If, however, the typical value of the sum is different from zero, changes in $x_i$ will not affect the sum. In this case, there is not Onsager's reaction term, which is precisely our situation here. Further intution behind this effect is nicely seen in, for instance, \cite{Shamir2000} in models of disordered magnets.}. Taking this into account, we obtain the following set of equations in the dense (sometimes rightfully called fully connected) limit:
\beeqn{
\hspace{-1cm}\Sigma(t)&=\frac{\alpha_{t}}{\alpha_{t-1}}\left(\frac{1}{\sigma(t)}-\frac{1}{J^2(t)}\frac{1}{\sigma(t)}\sigma(t+1)\frac{1}{\sigma(t)}\right)\,,\quad t=1,\ldots,M-1\,,\nonumber\\
\hspace{-1cm} \Sigma(M)&=-\frac{\alpha_{M}}{\alpha_{M-1}}\frac{J^2(M)}{1- J^2(M)\sigma(M)}\,\\
\hspace{-1cm}\sigma(1)&=\frac{1}{z+\Sigma(1)}\,,\quad \sigma(t+1)=\frac{1}{\Sigma(t+1)+\frac{1}{ J^2(t)\sigma(t)}}\,,\quad t=1,\ldots,M-1\,,
}
where we have used that $k_t=q_{t+1}\frac{\alpha_{t}}{\alpha_{t-1}}$ with $\alpha_{t-1}=N(t)/N(1)$ -notice that $\alpha_0=1$- and defined $\Sigma(t)\equiv\frac{\alpha_{t}}{\alpha_{t-1}}J^2(t)\gamma(t)$.\\
It is possible, as it was noticed in \cite{Burda2010b} by other means,  to show that $\sigma(1)$ must obey the polynomial equation (see its derivation in appendix \ref{ap:polynomial}):
\beeq{
\hspace{-1.5cm}P_{M+1}(v)=vz\,,\quad\quad P_{M+1}(v)=\prod_{s=1}^{M+1}\left(1+\frac{\alpha_M}{\alpha_{s-1}}v\right)\,,\quad\quad v=-\frac{1-z\sigma(1)}{\alpha_{M}}
\label{eq:polynomial}
}
cwhere we have rescaled $z/\mathcal{J}^2_M\to z$ and $\mathcal{J}^2_M\sigma(1)\to\sigma(1)$, with $\mathcal{J}_M=\prod_{s=1}^{M}J(s)$. Finally, and since in the dense limit $\Delta=\sigma(1)$, the spectral density is obtained using the formula:
\beeq{
\rho(\lambda)=\lim_{\epsilon\to 0^+}\frac{1}{\pi}{\rm Im}\left[\sigma(1)\right]\big|_{z=\lambda-i\epsilon}
\label{eq:sdfc}
}
The rest of this work consists in analysing the solutions coming from the polynomial equation \eref{eq:polynomial}. Some words are in order, though. It is evidently clear that solutions by radicals can be found for $M$ up to three, but even for $M=3$ the resulting expressions are rather unpleasant. It is however possible to consider solutions for a particular set of parameters for which the resulting expressions are much  better-looking, while still keeping their main properties somewhat intact. With this in mind, let us introduce some notation.  Recalling that  $X(1)\cdots X(M)$, we denote as $\bm{\alpha}=(\alpha_1,\ldots,\alpha_M)$ the standard choice of parameters  and  as $\bm{\alpha}^{(s)}=(1,\overset{s-1}{\ldots},1,\alpha,\ldots,\alpha)$ the choice of parameters to describe the situation in which the first $s-1$-matrices are square of unit size, the $s$-matrix is rectangular of relative size $1\times \alpha$ and the rest are square of size $\alpha\times \alpha$, that is:
\def\matriximg{%
  \begin{array}{ccc}
    &\cdots&\\
    \vdots&\ddots&\vdots\\
    &\cdots&
   \end{array}
}%
{\small \[\hspace{-3cm}
  {\scriptsize 1}\left\{\left(\vphantom{\matriximg}\right.\right.\kern-2\nulldelimiterspace
  \overbrace{\matriximg}^{{1}}\kern-\nulldelimiterspace\left.\vphantom{\matriximg}\right)\times
\left(\vphantom{\matriximg}\right.\kern-2\nulldelimiterspace
  \overbrace{\matriximg}^{{1}}\kern-\nulldelimiterspace\left.\vphantom{\matriximg}\right)\times\cdots  \times
\underbrace{\left(\vphantom{\matriximg}\right.\kern-2\nulldelimiterspace
  \overbrace{\matriximg}^{\alpha}\kern-\nulldelimiterspace\left.\vphantom{\matriximg}\right)}_{s-{\rm matrix}}\times
\left(\vphantom{\matriximg}\right.\kern-2\nulldelimiterspace
  \overbrace{\matriximg}^{\alpha}\kern-\nulldelimiterspace\left.\vphantom{\matriximg}\right)\times\cdots\times
\left(\vphantom{\matriximg}\right.\kern-2\nulldelimiterspace
  \overbrace{\matriximg}^{\alpha}\kern-\nulldelimiterspace\left.\vphantom{\matriximg}\right)
\]}

\section{Exact results by radicals for $M=1$, $2$ and $3$}
\label{m123}
\subsection{Case $M=1$}
For $M=1$, after solving the quadratic equation for $\sigma(1)$ and extracting from it the spectral density using \eref{eq:sdfc}, we obtain the following expression:
\beeqn{
\rho(\lambda)&=(1-\min(1,\alpha_1))\delta(\lambda)+\frac{1}{2\pi \lambda}\sqrt{(\lambda_{+}-\lambda)(\lambda-\lambda_{-})}\bm{1}_{\lambda\in[\lambda_{-},\lambda_{+}]}
}
with $\lambda_{\pm}=(1\pm\sqrt{\alpha_1})^2$. This is as expected the Marcenko-Pastur law. We would like to point out, as this will be used later on, that for small $\alpha_1$, the continuous part of the spectrum tends to a Wigner law after proper scaling. Indeed, for a small $\alpha_1$ we have that $\lambda_{\pm}=1\pm2\sqrt{\alpha_1}+\cdots$. If we define $\lambda=1+2\sqrt{\alpha_1}w$ we have that that the continuous part gives goes to $\alpha_1\frac{2}{\pi}\sqrt{1-w^2}$ for $w\in[-1,1]$.
\subsection{Case $M=2$}
\begin{figure}[h]
\begin{center}
\begin{picture}(320,320)
\put(-70,140){$\rho(\lambda)$}
\put(-70,310){$\rho(\lambda)$}
\put(155,140){$\rho(\lambda)$}
\put(155,310){$\rho(\lambda)$}
\put(150,-6){$\lambda$}
\put(390,-6){$\lambda$}
\put(150,170){$\lambda$}
\put(390,170){$\lambda$}
\put(-50,175){\includegraphics[width=7.5cm,height=5cm]{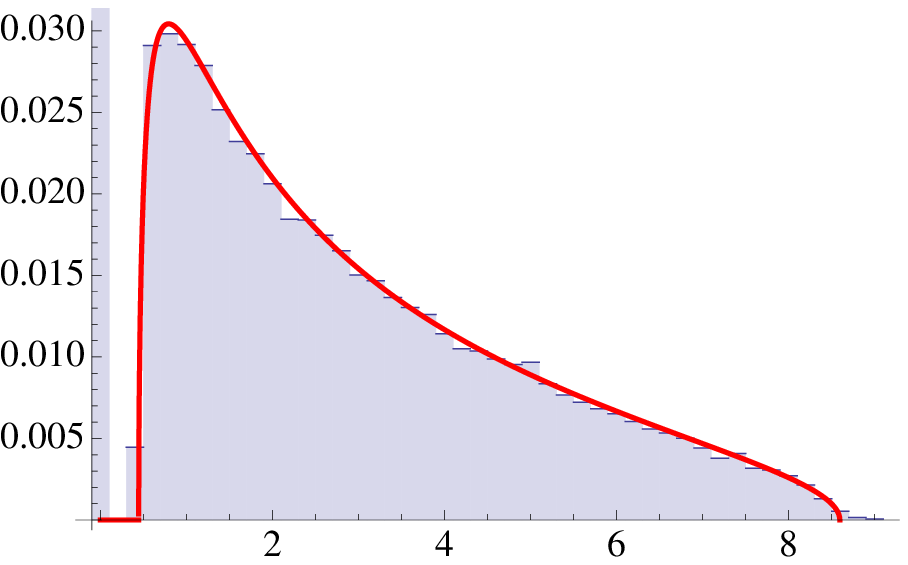}\quad\includegraphics[width=7.5cm,height=5cm]{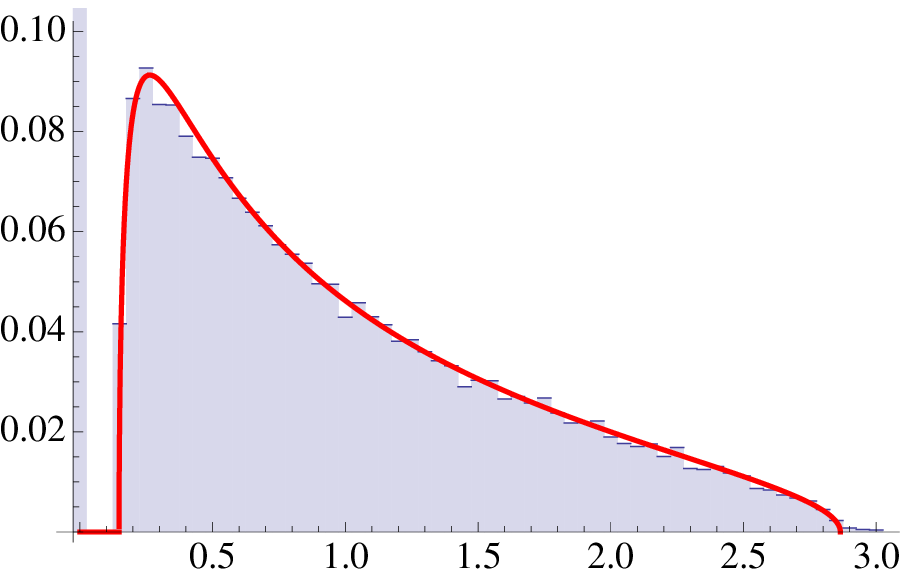}}
\put(-50,0){\includegraphics[width=7.5cm,height=5cm]{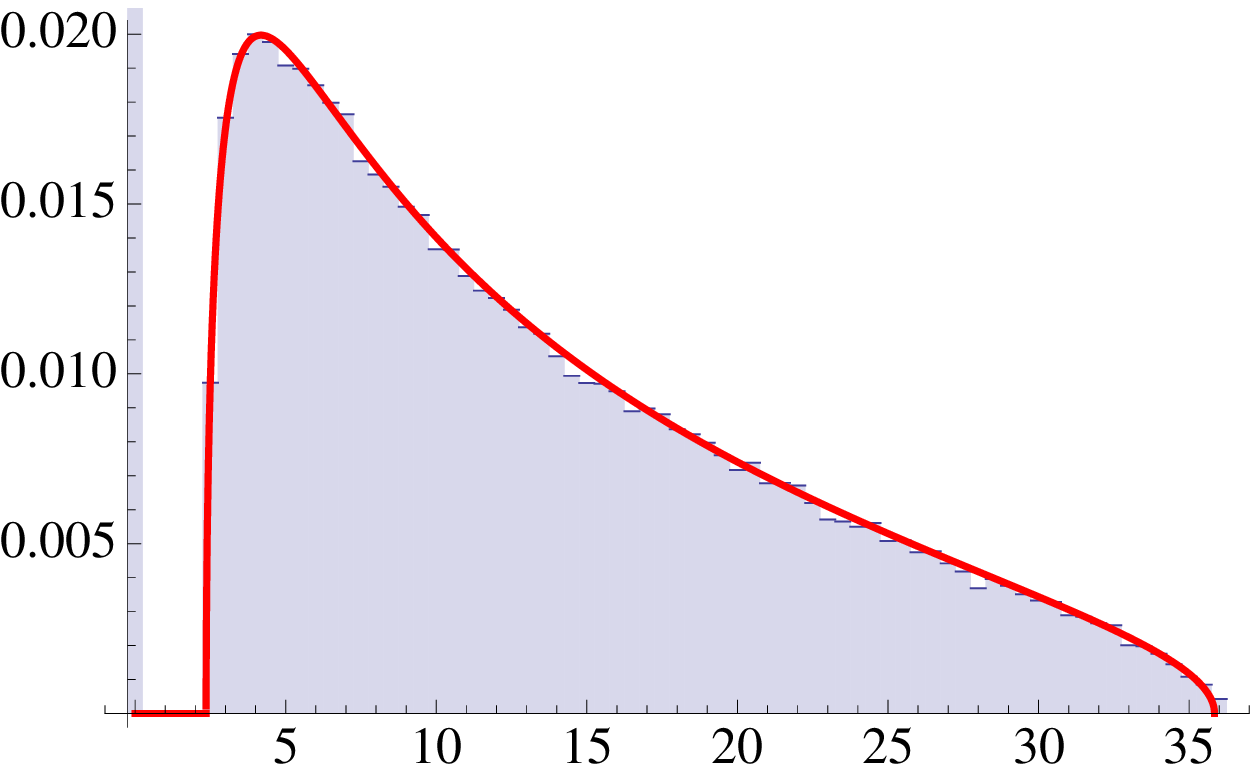}\quad\includegraphics[width=7.5cm,height=5cm]{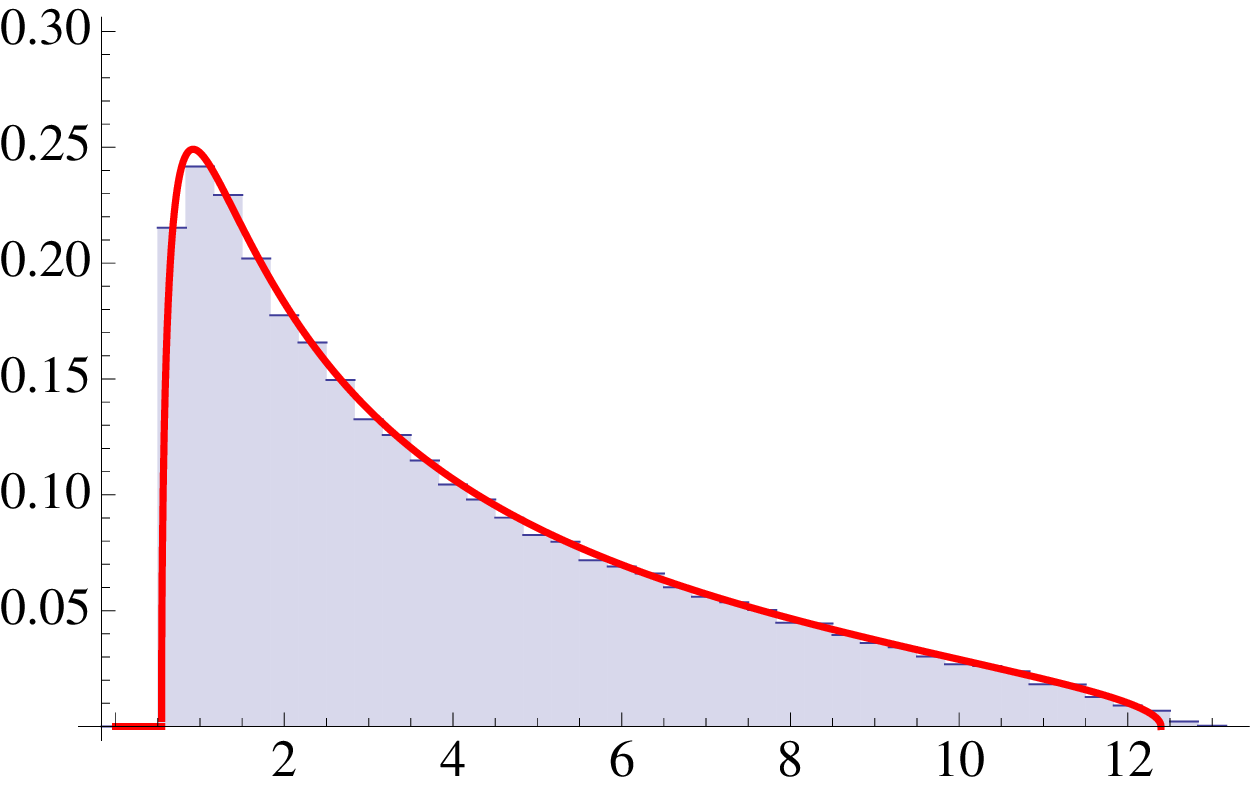}}
\end{picture}
\caption{Clockwise with notation $(\alpha_1,\alpha_2)$ we have: $(0.1,0,3)$ (for histogram we have $N(1)=1000$, $N(2)=100$ and $N(3)=300$  averaged over 1000 samples), $(0.3,0.1)$ (for histogram we have $N(1)=1000$, $N(2)=300$ and $N(3)=100$  averaged over 1000 samples), $(4,4)$ (with $N(1)=100$, $N(2)=400$ and $N(3)=400$d averaged over 1000 samples), and $(0.3,4)$ ($N(1)=500$, $N(2)=150$ and  $N(3)=2000$ averaged 1000 samples).}
\label{fig:m2}
\end{center}
\end{figure}

After analysing the cubic polynomial in detailed and looking for the best way to simplify the expressions (see appendix \ref{ap:casem2}) we find that the spectral density reads:
\beeqn{
\hspace{-1cm}\rho(\lambda)&=(1-\min(1,\alpha_1,\alpha_2))\delta(\lambda) \\
\hspace{-1cm}&+\frac{\sqrt{3}}{ 6\pi \lambda \sqrt[3]{2}}\Bigg(  \sqrt[3]{9\alpha_1\gamma(\lambda-\xi_0) + 6\sqrt{3\alpha_1^3(\lambda-\lambda_{-})(\lambda_{+,2}-\lambda)(\lambda-\lambda_{+,1})}} \\
\hspace{-1cm}&- \sqrt[3]{9\alpha_1\gamma(\lambda-\xi_0) - 6\sqrt{3\alpha_1^3(\lambda-\lambda_{-})(\lambda_{+,2}-\lambda)(\lambda-\lambda_{+,1})}}\Bigg)\bm{1}_{\lambda\in[\lambda_{+,1},\lambda_{+,2}]}
}
with
\beeqn{
\hspace{-1cm}\xi_0=-\frac{2(1+\alpha_1+\alpha_2)^3-9(\alpha_1(1+\alpha_1)+\alpha_2(1+\alpha_2)+\alpha_1\alpha_2(\alpha_1+\alpha_2))}{9\alpha_1(1+\alpha_1+\alpha_2)}
}
Here $\lambda_{-}$, $\lambda_{+,1}$, and $\lambda_{+,2}$ are the roots of a cubic the cubic polynomial reported in the appendix \ref{ap:casem2}. We do not report their explicit form for the general case,  as it suffices to point out that: (i) since $ (1 + \alpha_1 + \alpha_2)^3> 27\alpha_1\alpha_2$,  the three roots are real; (ii) for the cases $\alpha_1=1$ or $\alpha_2=1$ or $\alpha_1=\alpha_2$, the polynomial coefficient $d=0$, meaning that one root is zero. Any of these choices correspond precisely to the family of parameters mentioned above; (iii)  generally, there is one a non-positive root $\lambda_{-}$ and two non-negatives roots $\lambda_{+,1}$ and $\lambda_{+,2}$ with $\lambda_{+,2}>\lambda_{+,1}$.\\
In figure \ref{fig:m2} we have plotted the spectral density for four different values of the pair of parameters $(\alpha_1,\alpha_2)$ and compared them with numerical diagonalisation.

From the analysis of the roots and the expression of the continuous part of the spectral density we have that $\rho_{{cont}}(w,\alpha_1,\alpha_2)=(\alpha_1/\alpha_2)\rho_{{cont}}\left((\alpha_1/\alpha_2)w,\alpha_2,\alpha_1\right)$, scaling that, by simple eye inspection, can be numerically observed in figure \ref{fig:m2}.\\
Let us move on to analyse the cases for which the polynomial coefficient $d=0$.
\subsubsection{$\bm{\alpha}^{(1)}=(\alpha,\alpha)$ (first rectangular, second square)}
\begin{figure}[h]
\begin{center}
\begin{picture}(320,160)
\put(-75,140){$\rho(\lambda)$}
\put(150,140){$\rho(\lambda)$}
\put(150,-6){$\lambda$}
\put(390,-6){$\lambda$}
\put(-50,0){\includegraphics[width=7.5cm,height=5cm]{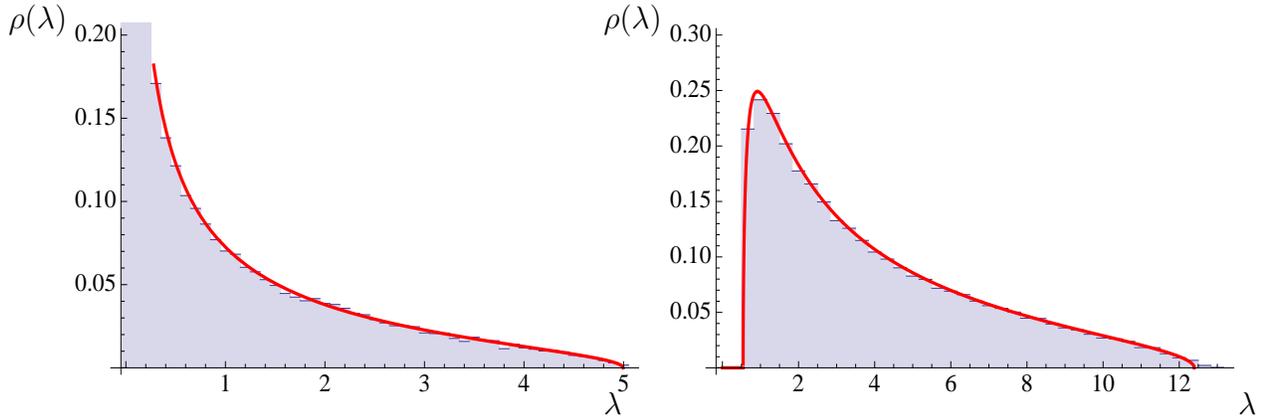}\quad\includegraphics[width=7.5cm,height=5cm]{figures/His100_400_400_1000.eps}}
\end{picture}
\caption{Results for $\bm{\alpha}^{(1)}=(\alpha,\alpha)$ with $\alpha=3/10$ (histograms have been performed for $N(1)=500$, $N(2)=N(3)=150$ averaged over 1000 samples) at the left where there is also a Dirac delta contribution at zero and for $\alpha=4$ (histograms have been performed for $N(1)=100$, $N(2)=N(3)=400$ averaged over 1000 samples) at the right plot.}
\label{fig:m2a1}
\end{center}
\end{figure}

This case is interesting as it yields a different scenario to the one observed for the Marcenko-Pastur law, namely,  having no gap and Dirac delta contribution at zero. Indeed, the spectral density  here reads:
\beeqn{
\hspace{-1cm}\rho(\lambda)&= \left(1-\min(1,\alpha)\right)\delta(\lambda) \\
\hspace{-1cm}&+\frac{\sqrt{3}}{ 6\pi \lambda \sqrt[3]{2}}\Bigg(  \sqrt[3]{9\alpha(1+2\alpha)(\lambda-\xi_0) + 6\sqrt{3\alpha^3(\lambda-\lambda_{-})\lambda(\lambda_{+}-\lambda)}}\\
\hspace{-1cm}&- \sqrt[3]{9\alpha(1+2\alpha)(\lambda-\xi_0) - 6\sqrt{3\alpha^3(\lambda-\lambda_{-})\lambda(\lambda_{+}-\lambda)}}\Bigg)\bm{1}_{\lambda\in[\lambda_{-}\Theta(\alpha-1),\lambda_{+}]}
}
with
\beeqn{
\lambda_{\pm}&=\frac{-1+20\alpha+8\alpha^2\pm(1+8\alpha)^{3/2}}{8\alpha}\,,\quad \xi_0=\frac{2(-1+\alpha)^3}{9\alpha(1+2\alpha)}\,.
}
Notice that $\lambda_-\leq0$ for $\alpha\in(0,1]$, while $\lambda_{-}$ is positive for $\alpha\in(1,\infty)$. At the same time  $\lambda_{+}\geq 4$ from $\alpha\in[0,\infty)$. Besides $\lambda_{-}\simeq -\frac{1}{4\alpha}$ as $\alpha\to0$.  The attentive reader surely have realised that for $\alpha\in(0,1)$ there is both and a continuous density at zero as well as a Dirac delta contribution, while for $\alpha>1$ the Dirac delta disappears and a gap appears in the continuous part (see figure \ref{fig:m2a1}).\\
For small $\alpha$ we can see better the combined contribution at zero of the continuous and isolated part. Moreover, the continuous part of the spectrum disappears with a factor $\alpha$ while keeping a well-defined shape reminiscent of the Marcenko-Pastur law for $\alpha=1$, viz.
\beeqn{
\rho(\lambda)&= \left(1-\alpha\right)\delta(\lambda)+\frac{\alpha}{2\pi}\sqrt{\frac{4-\lambda}{\lambda}}\bm{1}_{\lambda\in[0,4]}+\mathcal{O}(\alpha^2)
}
The curious reader would have noticed that the first term in $\alpha$ is such that the spectral density is normalised to one. One wonders what happens to the normalisation of higher-order terms in the expansion: first of all the functions of higher-order terms are not densities (they can generally have positive and negative values); secondly, apart of the $\alpha^2$ correction whose integral is zero, the integral of the rest of the terms diverges term by term.

\subsubsection{$\bm{\alpha}^{(2)}=(1,\alpha)$ (first matrix square, second rectangular)}
\begin{figure}[h]
\begin{center}
\begin{picture}(320,160)
\put(-75,140){$\rho(\lambda)$}
\put(150,140){$\rho(\lambda)$}
\put(150,-6){$\lambda$}
\put(390,-6){$\lambda$}
\put(-50,0){\includegraphics[width=7.5cm,height=5cm]{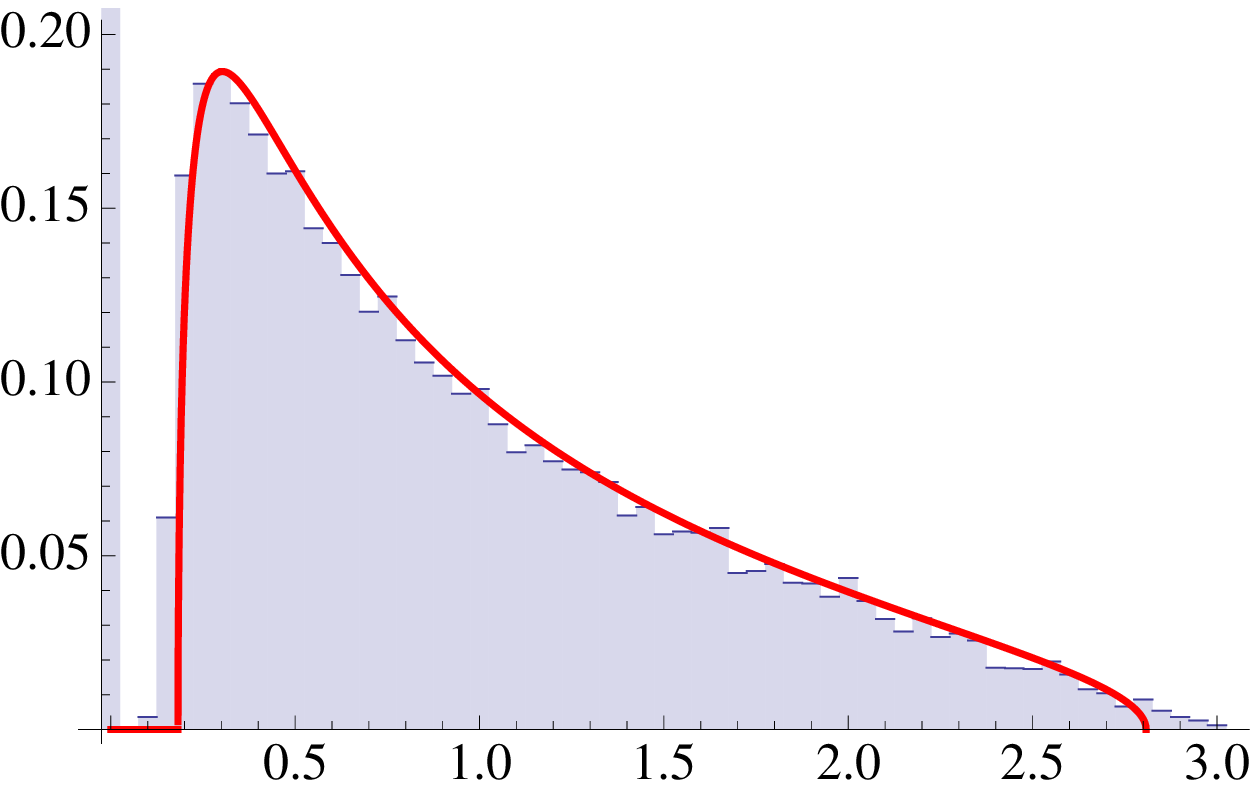}\quad\includegraphics[width=7.5cm,height=5cm]{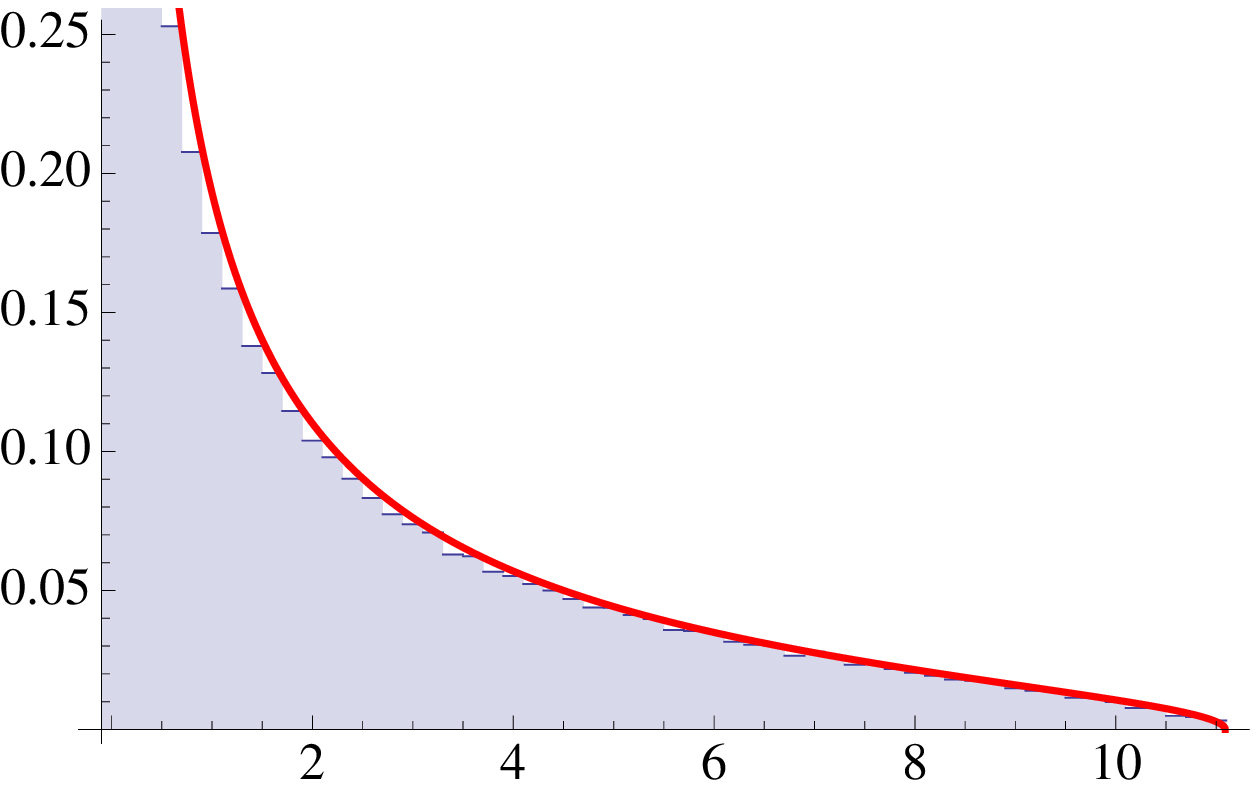}}
\end{picture}
\caption{Case $\bm{\alpha}^{(2)}=(1,\alpha)$ for two values of $\alpha=2/10$ (for the histograms we took $N(1)=100$, $N(2)=100$ and $N(3)=20$ averaed over 1000 samples) and $\alpha=2$ (for the histograms we took $N(1)=100$, $N(2)=100$ and $N(3)=200$ averaged over 1000 samples)}
\label{fig:m2anothercase}
\end{center}
\end{figure}

In this case, the spectral density becomes:
\beeqn{
\hspace{-1cm}\rho(\lambda)&=(1-\min(1,\alpha))\delta(\lambda)\\
\hspace{-1cm}&+\frac{\sqrt{3}}{ 6\pi \lambda \sqrt[3]{2}}\Bigg(  \sqrt[3]{9(2+\alpha)(\lambda-\xi_0) + 6\sqrt{3(\lambda-\lambda_{-})\lambda(\lambda_{+}-\lambda)}}\\
\hspace{-1cm}&- \sqrt[3]{9(2+\alpha)(\lambda-\xi_0) - 6\sqrt{3(\lambda-\lambda_{-})\lambda(\lambda_{+}-\lambda)}}\Bigg)\bm{1}_{\lambda\in[\lambda_{-}\Theta(1-\alpha),\lambda_{+}]}\nonumber
}
with
\beeqn{
\lambda_{\pm}&=\frac{1}{8}(8+20\alpha-\alpha^2\pm\sqrt{\alpha}(8+\alpha)^{3/2})\,,\quad  \xi_0=-\frac{2(-1+\alpha)^3}{9(2+\alpha)}
}
We have that $\lambda_+\geq \lambda_-$ the equality only resulting when $\alpha=0$ in which case $\lambda_-=\lambda_+=1$. Now, we have that $\lambda_{+}$ is always positive while  $\lambda_-$ is positive for $\alpha\leq 1$ and then $\lambda_-<0$ for $\alpha\in(1,\infty)$. This implies that there will be a gap for $\alpha\in[0,1]$ with Dirac delta, while for $\alpha>1$ there is no gap and no Dirac delta, which again differs from the well-known properties of the  Marcenko-Pastur law. In the figure \ref{fig:m2anothercase} we plot the two cases.

It is interesting to see what happens for small $\alpha$, since this asymptotic behaviour, as we will see later on, depends strongly on the position of the rectangular matrix. We note that for small $\alpha$ the endpoints have the expansion $\lambda_{-}=1-2\sqrt{2}\sqrt{\alpha}+\cdots$, $\lambda_{+}=1+2\sqrt{2}\sqrt{\alpha}+\cdots$. This implies that the continuous part of the spectrum shrinks to zero changing its shape, unlike the previous case for $\bm{\alpha}^{(1)}$. To obtain the limiting vanishing shape, we do the change of variables $\lambda=2\sqrt{2 \alpha}w +1$ to obtain that  $\rho_{cont}\to \alpha\frac{2}{\pi}\sqrt{1-w^2}$,that is, the Wigner law.\\
Let us finish by noticing that, similarly to the symmetry of the Marcenko-Pastur law relating $\alpha\leftrightarrow 1/\alpha$, we have here that the spectral densities corresponding to the two choices of parameters are related by a transformation $(\lambda,\alpha)\to (\lambda/\alpha,1/\alpha)$.

\subsection{Case $M=3$}
Finally, for $M=3$ the spectral density is given by the following expression
\beeqn{
\hspace{-1cm}\rho(\lambda)&= (1-\min(1,\alpha_1,\alpha_2,\alpha_3))\delta(\lambda)\\
\hspace{-1cm}&+\frac{1}{2\pi \lambda}\sqrt{4\mathcal{S}^2(\lambda) + \frac{\gamma^2-4\gamma_2}{4} + \frac{1}{8}\frac{Q_1(\lambda)}{\mathcal{S}(\lambda)}}\bm{1}_{\lambda\in[\lambda_{+,1},\lambda_{+,2}]}
}
with
\beeqn{
\hspace{-1cm}\mathcal{S}(\lambda)&=\frac{1}{2}\sqrt{- \frac{\gamma^2-4\gamma_2}{12}+\frac{1}{3}\left(\sqrt[3]{\frac{P_2(\lambda) + \sqrt{P_4(\lambda)}}{2}} + \frac{P_1(\lambda)}{\sqrt[3]{\frac{P_2(\lambda) + \sqrt{P_4(\lambda)}}{2}}}\right)}
}
where the derivation and definitions of the polynomials involved can be found in appendix \ref{ap:casem3}. Here $\lambda_{-}$ and $\lambda_{+}$ are roots from solving $4\mathcal{S}^2(\lambda) + \frac{\gamma^2-4\gamma_2}{4} + \frac{1}{8}\frac{Q_1(\lambda)}{\mathcal{S}(\lambda)}=0$. These two roots  seems to be related to the zeros of polynomial $P_4(\lambda)$, albeit in a complicated manner at least for the lower endpoint.\\
As the general solution is fairly involved mathematically, we look for a choice of parameters which simplify the zeros of $P_4$. This choice turns out to be the ones we introduced above.
\subsubsection{ $\bm{\alpha}^{(1)}=(\alpha,\alpha,\alpha)$ (recangular-square-square)}
\begin{figure}[h]
\begin{center}
\begin{picture}(320,160)
\put(-70,140){$\rho(\lambda)$}
\put(155,140){$\rho(\lambda)$}
\put(150,-6){$\lambda$}
\put(390,-6){$\lambda$}
\put(-50,0){\includegraphics[width=7.5cm,height=5cm]{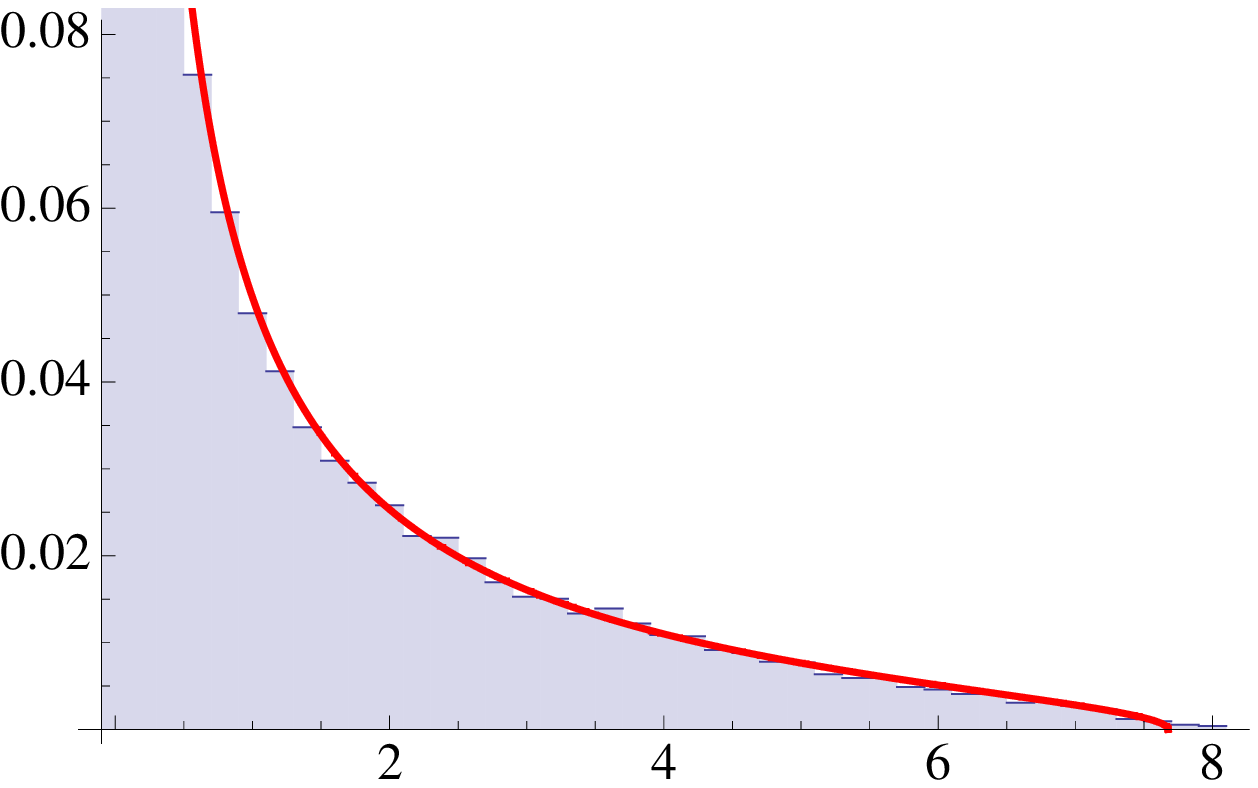}\quad\includegraphics[width=7.5cm,height=5cm]{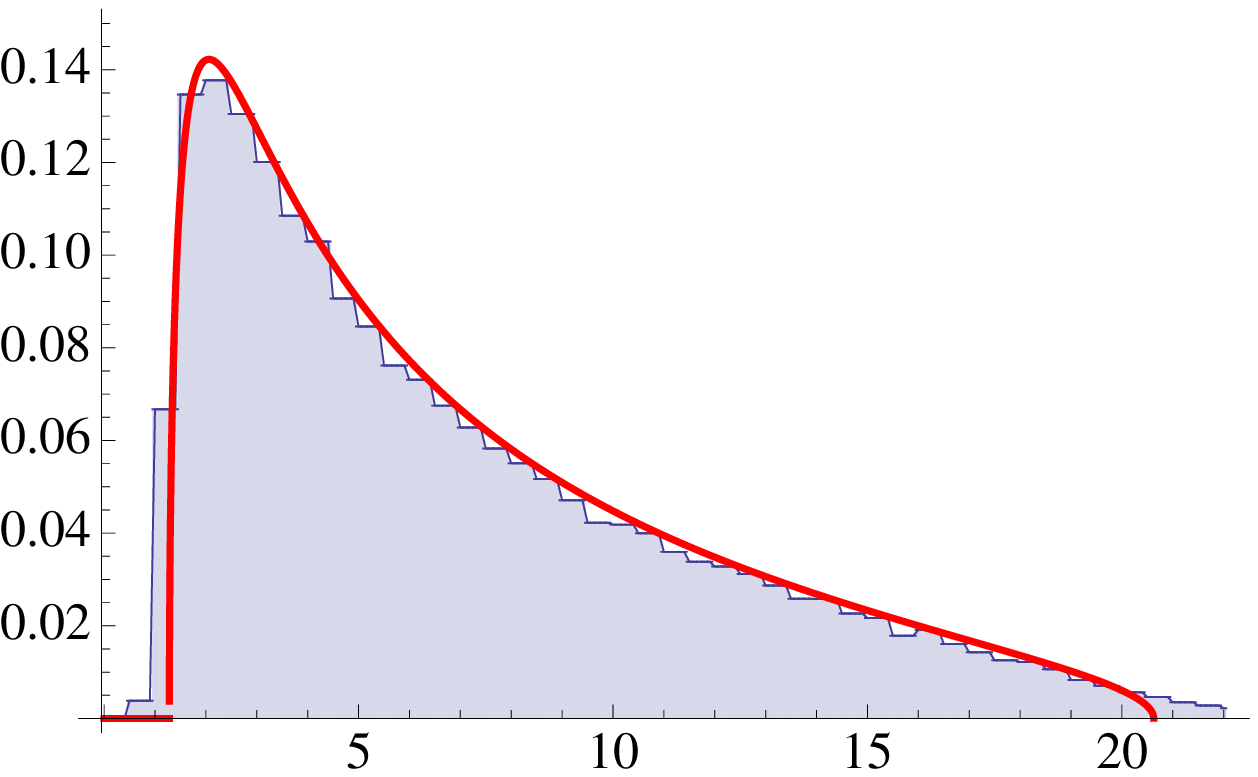}}
\end{picture}
\caption{Results for  $\bm{\alpha}^{(1)}=(\alpha,\alpha,\alpha)$. Left: $\alpha=3/10$ and there is a Dirac delta not visible as it mixes up with the continuous part (for the histograms we have chosen $N(1)=500$ and $N(2)=N(3)=N(4)=150$  averaged over 1000 samples). Right: $\alpha=7$ (for the histograms we have chosen $N(1)=10$ and $N(2)=N(3)=N(4)=70$ averaged over 1000 samples).}
\label{fig:m3cas1}
\end{center}
\end{figure}

Particularising for this case we have the following spectral density
\beeqn{
\hspace{-2.5cm}\rho(\lambda)&=(1-\min(1,\alpha))\delta(\lambda)+\frac{1}{2\pi \lambda}\sqrt{4\mathcal{S}^2(\lambda) - \frac{3(1-\alpha)^2}{4} - \frac{\alpha^2\left(\lambda+\frac{(-1+\alpha)^3}{8\alpha^2}\right)}{\mathcal{S}(\lambda)}}\bm{1}_{\lambda\in[\Theta(\alpha-1)\lambda_{-},\lambda_{+}]}
}
with
\beeqn{
\hspace{-1cm}\mathcal{S}(\lambda)&=\frac{1}{2}\Bigg( \frac{(1-\alpha)^2}{4}+\left(\frac{\alpha^4 \lambda}{2}\right)^{1/3}\sqrt[3]{\lambda+\frac{(1-\alpha)^2}{\alpha} + \sqrt{(\lambda-\lambda_{-})(\lambda-\lambda_{+})}} \\
\hspace{-1cm}&+ \frac{\alpha^{2/3}(1+3\alpha)\lambda}{3\left(\frac{ \lambda}{2}\right)^{1/3}\sqrt[3]{\lambda+\frac{(1-\alpha)^2}{\alpha} + \sqrt{(\lambda-\lambda_{-})(\lambda-\lambda_{+})}}}\Bigg)^{1/2}
}
and
\beeqn{
\lambda_{\pm}&=\frac{2 \pm 2 \sqrt{(1 + 3 \alpha)^3 (-1 + 9 \alpha)^2} + 9 \alpha (-1 + 3 \alpha (4 + \alpha))}{27 \alpha^2}
}
In the figure \ref{fig:m3cas1} we show some plots and their comparison with numerical diagonalisation.

The astute reader has surely already realised what is to be expected when considering  small $\alpha$, by simply looking at the number of square matrices of size $\alpha$. Indeed for small $\alpha$ we obtain
\beeqn{
\hspace{-2cm}\rho(\lambda)&=(1-\alpha)\delta(\lambda)\\
\hspace{-2cm}&+\frac{\alpha}{2^{4/3}3^{1/6}\pi \lambda}\left[\left(9\lambda+\sqrt{3}\lambda\sqrt{27-4\lambda}\right)^{1/3}-\left(9\lambda-\sqrt{3}\lambda\sqrt{27-4\lambda}\right)^{1/3}\right]\bm{1}_{\lambda\in[0,\frac{27}{4}]}\\
\hspace{-2cm}&+\mathcal{O}(\alpha^2)\,,
}
that is, the continous part keeps a definite shape -the one corresponding to the spectral density of $M=2$ for $\alpha_1=\alpha_2=1$- while vanishing with an overall factor $\alpha$.
\subsection{Case $\bm{\alpha}^{(2)}=(1,\alpha,\alpha)$ (square-rectangular-square)}
Particularising the formulas in appendix \ref{ap:casem3} for this case we obtain
\beeqn{
\hspace{-1.5cm}\rho(\lambda)&=(1-\min(1,\alpha))\delta(\lambda)+\frac{1}{2\pi \lambda}\sqrt{4\mathcal{S}^2(\lambda) -(1-\alpha)^2 -\frac{\alpha \lambda}{\mathcal{S}(\lambda)}}\bm{1}_{\lambda\in[0,\lambda_{+}]}
}
with
\beeqn{
\hspace{-2cm}\mathcal{S}(\lambda)&=\frac{1}{2\sqrt{3}}\Bigg((1-\alpha)^2+3\alpha^{2/3}\sqrt[3]{\frac{ (\lambda-p_-)(\lambda-p_+) + \lambda\sqrt{ (\lambda-\lambda_-)(\lambda-\lambda_{+})}}{2}}\\
\hspace{-2cm}&+ \frac{(1-\alpha)^4+6\alpha(1+\alpha)\lambda}{3\alpha^{2/3}\sqrt[3]{\frac{ (\lambda-p_-)(\lambda-p_+) + \lambda\sqrt{(\lambda-\lambda_-)(\lambda-\lambda_{+})}}{2}}}\Bigg)^{1/2}
}
and
\beeqn{
\hspace{-1cm} \lambda_{\pm}&=\frac{2 \left(\alpha (33-(\alpha-33) \alpha)\pm\sqrt{(\alpha (\alpha+14)+1)^3}-1\right)}{27 \alpha}\,,\quad \\
\hspace{-1cm}p_{\pm}&=\frac{-3 (\alpha-1)^2 \alpha (\alpha+1)\pm\sqrt{3} \sqrt{(\alpha-1)^4 \alpha^2 (\alpha (\alpha+10)+1)}}{9 \alpha^2}\,.
}
This case yields a situation where there is \textit{no gap for any value of $\alpha$}. Actually, as we will see later this \textit{no gap property}, it is generally observed for the family of parameter $\bm{\alpha}^{(s)}$ for $s=2,\ldots,M-1$.\\
For small $\alpha$ we again see that the continuous part of the spectral density keeps a well-defined shape, viz
\beeqn{
\rho(\lambda)=(1-\alpha)\delta(\lambda)+\alpha\frac{1}{4\pi}\sqrt{\frac{4-\lambda}{\lambda}}+\cdots
}
which corresponds, as expected, to the spectral density for $M=1$ and $\alpha_1=1$.
\subsubsection{Case $\bm{\alpha}^{(3)}=(1,1,\alpha)$ (square-square-rectangular)}
\begin{figure}[h]
\begin{center}
\begin{picture}(320,160)
\put(-75,140){$\rho(\lambda)$}
\put(155,140){$\rho(\lambda)$}
\put(150,-6){$\lambda$}
\put(390,-6){$\lambda$}
\put(-50,0){\includegraphics[width=7.5cm, height=5cm]{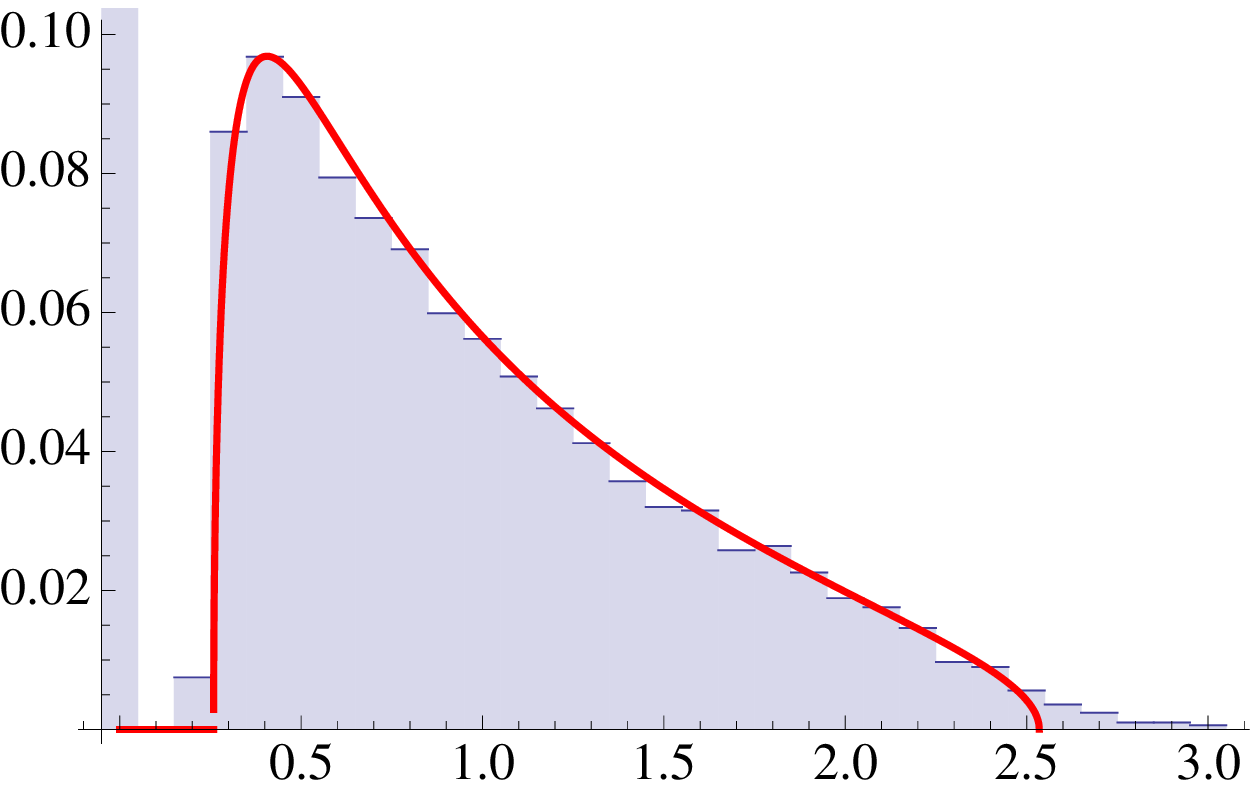}\quad\includegraphics[width=7.5cm, height=5cm]{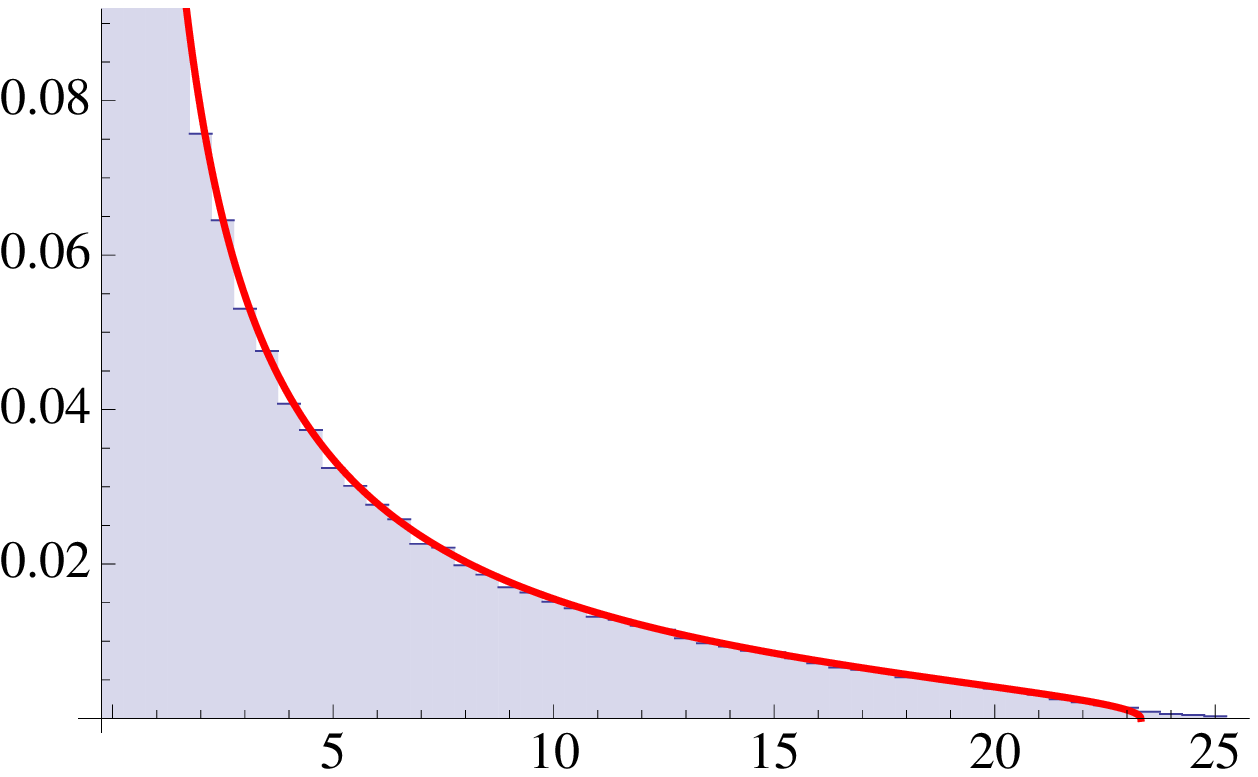}}
\end{picture}
\caption{Case $\bm{\alpha}^{(3)}=(1,1,\alpha)$ for $\alpha=1/10$ (for the histograms we have used $N(1)=N(2)=N(3)=100$ and $N(4)=10$ averaged over 1000 samples) with Dirac delta and $\alpha=3$ (for the histograms we have used $N(1)=N(2)=N(3)=100$ and $N(4)=300$ averaged over  1000 samples).}
\label{fig:m3case3}
\end{center}
\end{figure}

For this final choice of parameter we obtain
\beeq{
\hspace{-1cm}\rho(\lambda)&=(1-\min(1,\alpha))\delta(\lambda)\nonumber\\
\hspace{-1cm}&+\frac{1}{2\pi \lambda}\sqrt{4\mathcal{S}^2(\lambda) - \frac{3(1-\alpha)^2}{4} -\frac{1}{8}\frac{8\lambda+(1-\alpha)^3}{\mathcal{S}(\lambda)}}\bm{1}_{\lambda\in[\lambda_{-}\Theta(1-\alpha),\lambda_{+}]}
}
with
\beeqn{
\hspace{-1cm}\mathcal{S}(\lambda)&=\frac{1}{2}\Bigg( \frac{(1-\alpha)^2}{4}+\lambda^{1/3}\sqrt[3]{\frac{\lambda+(1-\alpha)^2 + \sqrt{(\lambda-\lambda_-)(\lambda-\lambda_+)}}{2}} \\
\hspace{-1cm}&+ \frac{(3+\alpha)\lambda^{2/3}}{3\sqrt[3]{\frac{\lambda+(1-\alpha)^2 + \sqrt{(\lambda-\lambda_-)(\lambda-\lambda_+)}}{2}}}\Bigg)^{1/2}
}
and
\beeqn{
\lambda_{\pm}&=\frac{1}{27} \left(\alpha (\alpha (2 \alpha-9)+108)\pm2 \sqrt{(\alpha-9)^2 \alpha (\alpha+3)^3}+27\right)\,.
}
Here somewhat the situation is reverse, namely, there is a gap for $\alpha <1$ and no gap for $\alpha\geq 1$. In figure \ref{fig:m3case3} we show some plots of our theoretical predictions and comparions with numerical diagonalisation.

\noindent Finally, for small $\alpha$ we find again that the continuous part is proportional to the Wigner law after doing the change of variables $\lambda= 2\sqrt{3\alpha}w+1$.\\
We finish by pointing out that the cases $\bm{\alpha}^{(1)}$ and $\bm{\alpha}^{(3)}$ are related by $(\lambda,\alpha)\to(\lambda/\alpha,1/\alpha)$.
\section{Three observations and some exact results for general $M$}
\label{generalm}

\subsection{Three numerical observations}
Based on the detailed derivations we have done for $M=1$, $2$ and $3$ and on some, admittedly rather weak, numerical experiments for larger values of $M$, we would like to point out three observations that seem to be correct.
Our first observation is:
\begin{prop}[The weight of the Dirac delta]
For any $M$ the weight associated with the Dirac delta is
\beeqn{
\omega(\bm{\alpha})=1-\min(1,\alpha_1,\ldots,\alpha_M)\,.
} 
\end{prop}
This is obviously true for $M=1$, $2$, and $3$ and seems to be true numerically for $M\geq 4$. However it is quite challenging to rigorously check this numerically for two reasons: firstly, for any value of $M$ (let us say moderately big) the error asociated with the numerical diagonalisation will not provide exactly zero eigenvalues but instead very small ones. If there were to be a gap between the zero -or very small- eigenvalues to the continuous spectrum, an estimate of the weight is certainly possible. However, in many cases there is either no gap or the gap is very small (in the sense that is is numerically indistinguisable as no having gap). We then run into the situation that the continuous part is mixed with the Dirac delta, not a pleasant situation to have numerically. A way around this is to calculate the cumulative probability function and then look at its derivative at zero. However, to have a good estimate for its derivative, the number of samples and the size of matrices must be large enough.\\

\noindent The second observation has to do with the lack of gap (but not necessarily of Dirac delta) when the ensemble of matrices is such than the first and the last matrices in $X(1)\cdots X(M)$ are square for the family of parameters $\bm{\alpha}^{(s)}$ :
\begin{prop}[No gap]
The continous part of  spectral density for the family of parameters $\bm{\alpha}^{(s)}$ for $s=2,\ldots,M-1$ has no gap in the spectral density for any value of the parameter $\alpha$.
\end{prop}
Again we would like to check somewhat carefully this observation numerically, but we find ourselves again in a difficult situation trying to disentangle the contribution of the Dirac delta from the continuous part.\\

\noindent The final observation has to do with the defined shape the spectral density takes for small $\alpha$ for the family of spectral densities with parameters $\bm{\alpha}^{(s)}$:
\begin{prop}[Asymptotic shape]
Consider the spectral density $\rho_M(\lambda|\bm{\alpha}^{(s)})$ for $s=1,\ldots,M$. Then for small $\alpha$ and for $s=1,\dots, M-1$ one finds that
 \beeqn{
\rho_{M}(\lambda|\bm{\alpha}^{(s)})=(1-\alpha)\delta(w)+\alpha\rho_{M-s}(\lambda|\{\alpha_i=1\}_{i=1}^{M-s})+\mathcal{O}(\alpha^2)
} 
while for $s=M$ one finds that the asymptotic behaviour of the continous part is given by the Wigner semi-circle law $\frac{2}{\pi}\sqrt{1-w^2}$ after rescaling to the new variable  $\lambda= 2\sqrt{M\alpha}w +1$.
\end{prop}
As we have seen this is certainly true for the exact cases we have discussed before. For $M\geq 4$ we would like to check this at least numerically. However we run into another impasse: to avoid finite size effects we must work  with large matrix sizes with small ratio $\alpha$. We could of course achieve small values of $\alpha$ using small matrices but then the finite size effects -not captured in this approach- are too significant to check carefully this observation.\\
A mathematically grounded explanation of these observations can be found in \cite{Akemann2013,Ipsen2013}.
\subsection{Endpoints of the continous part of the spectrum}
Starting from the polynomial equation \eref{eq:polynomial}, it is possible to derive equations for the endpoints of the support of the spectral density (this simple trick can also be found in \cite{Akemann2013}). Indeed, doing the derivative of $v$ with respect to $z$ and noticing that $dv/dz$ is not wel-defined at the endpoints, we obtain the following condition for $v(z_{{B}})$ at the boundary:
\beeq{
\frac{\alpha_{M} v(z_{{B}})}{1+\alpha_M v(z_{{B}})}+\sum_{\ell=1}^{M}\frac{\alpha_{M} v(z_{{B}})}{\alpha_{\ell}+\alpha_M v(z_{{B}})}=1\,.
\label{eq:boundaries}
}
We obviously cannot solve exactly this equation for any set of parameters $\bm{\alpha}$, but we can for the family of parameters used before.
\subsubsection{Square matrices}
Here $\bm{\alpha}=(1,\ldots,1)$ and the eq. \eref{eq:boundaries} gives $v(z_{{B}})=1/M$, which after inserting it in the polynomial equation \eref{eq:polynomial} yields the upper endpoint
\beeq{
\lambda_{+}=M\left(1+\frac{1}{M}\right)^{M+1}\,.
\label{eq:bsc}
}
Eq. \eref{eq:bsc} agrees with our previous ones for $M\leq 3$. We know that the lower endpoint is zero -as there is no gap- but we are not able to get $\lambda_{-}$ by starting from all square matrices, as we will see below.
\subsubsection{Family of parameters $\bm{\alpha}^{(s)}$}
For this family of parameters, the equation \eref{eq:boundaries} becomes
\beeqn{
s\frac{\alpha v(z_{{B}})}{1+\alpha v(z_{{B}})}+(M-s+1)\frac{ v(z_{{B}})}{1+v(z_{{B}})}=1\,.
}
Solving it and plugging the solution back into the the poylnomial equation, provides the following result:
\beeqn{
\hspace{-1cm}z_{\tau,{B}}(\alpha,M,s)&=\frac{1}{\Delta_{\tau}(\alpha,M,s)} \left(1+\alpha \Delta_{\tau}(\alpha,M,s)\right)^{s} \left(1+\Delta_{\tau}(\alpha,M,s)\right)^{M-s+1}\,,
}
for $\tau\in\{-1,1\}$ and for $s=1,\ldots,M$, and where we have defined
\beeqn{
\hspace{-1cm}\Delta_{\tau}(\alpha,M,s)=\frac{-\alpha (s-1)-M+s+\tau\sqrt{((\alpha-1) (s-1)+M-1)^2+4 \alpha M}}{2 \alpha M}\,.
}
A similar formula can be found in \cite{Akemann2013}. One should first note that for $\alpha=1$ we recover back eq. \eref{eq:bsc} plus the lower endpoint at zero. In the general case, however, $z_{\tau,{B}}(\alpha,M,s)$ gives the upper endpoint ($\tau=1$) correctly, while for the lower endpoint we must tweak the formula \footnote{This is what we also found in our analysis of $M=3$, in which the endpoints of spectral density are related to the roots of $P_4$. Actually, one can see that for $M=3$, $z_{\tau,{B}}(\alpha,M,s)$ provides two of the roots of $P_4$.}. The correct lower and upper endpoints turn out to be:
\beeqn{
\hspace{-2.5cm}\lambda_{-}(\alpha,M,s)&=z_{-,{B}}(\alpha,M,s)\left[\delta_{s,1}\Theta(\alpha-1)+\delta_{s,M}\Theta(1-\alpha)\right]\,,~ \lambda_{+}(\alpha,M,s)=z_{+,{B}}(\alpha,M,s)\,.
}
\subsubsection{Exact expression of the spectral density for square matrices}
\begin{figure}[h]
\begin{center}
\begin{picture}(320,320)
\put(-70,140){$\rho(\lambda)$}
\put(-70,310){$\rho(\lambda)$}
\put(155,310){$\rho(\lambda)$}
\put(155,140){$\bracket{\lambda_{+}}$}
\put(150,-6){$\lambda$}
\put(390,-6){$M$}
\put(150,170){$\lambda$}
\put(390,170){$\lambda$}
\put(-50,0){\includegraphics[width=7.5cm,height=5cm]{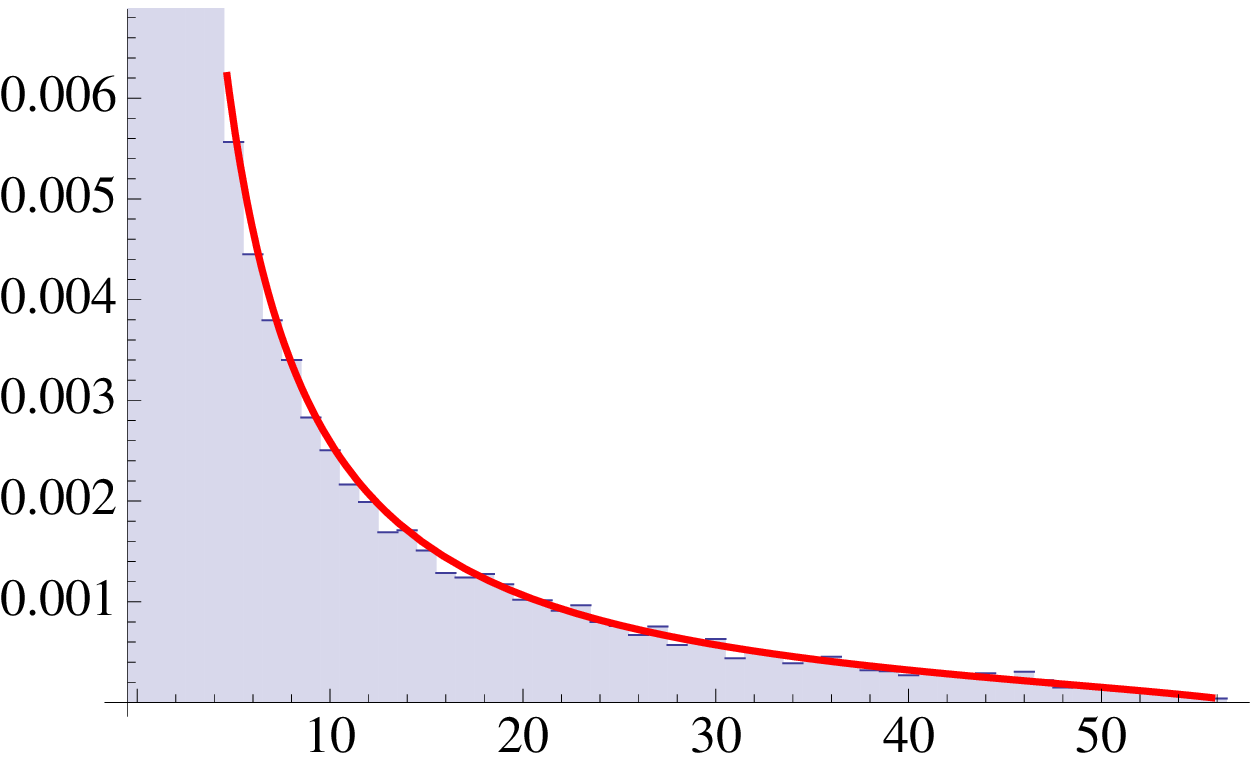}\quad\includegraphics[width=7.5cm,height=5cm]{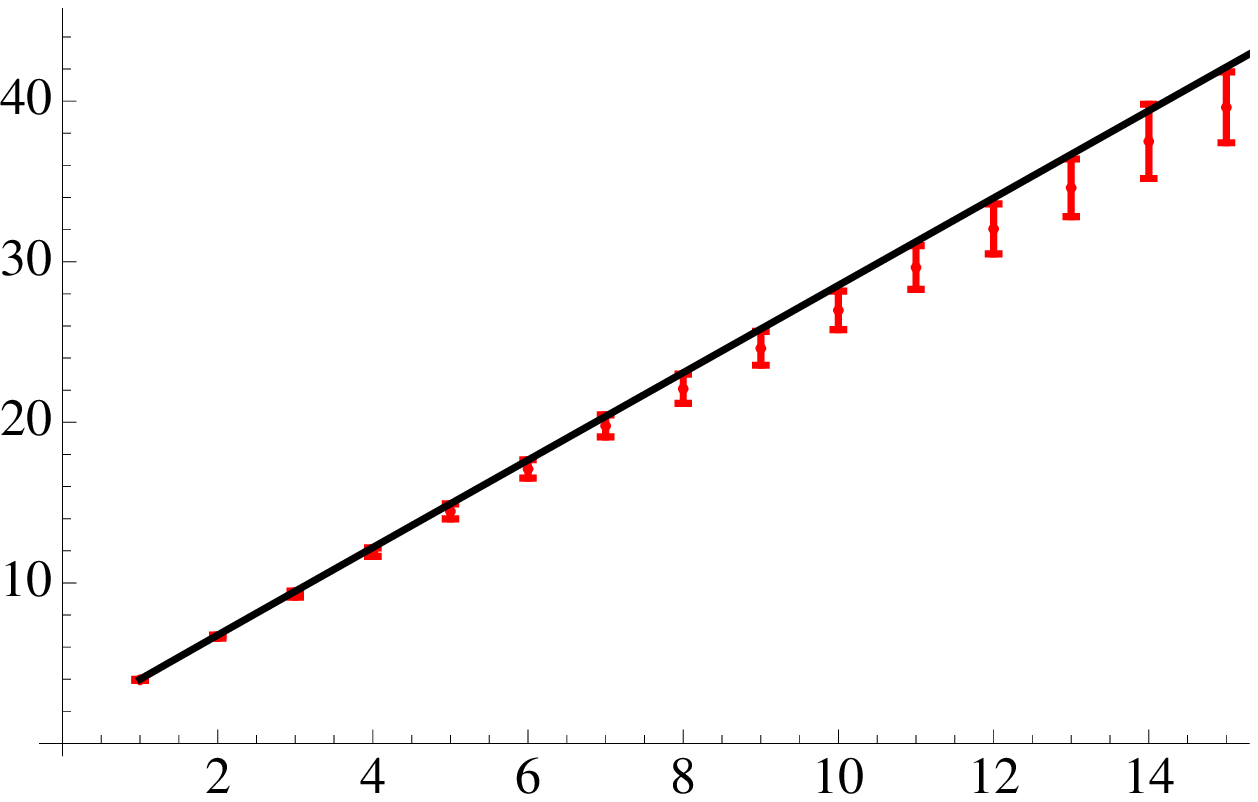}}
\put(-50,175){\includegraphics[width=7.5cm,height=5cm]{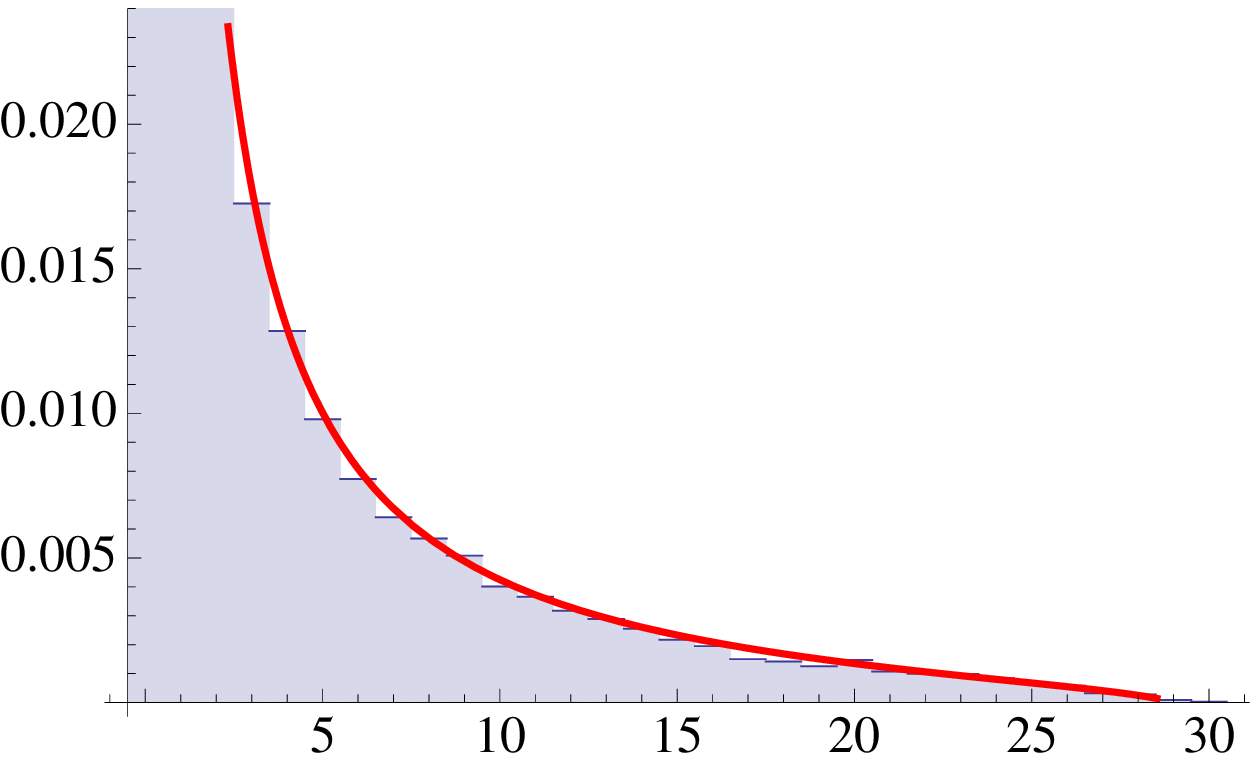}\quad\includegraphics[width=7.5cm,height=5cm]{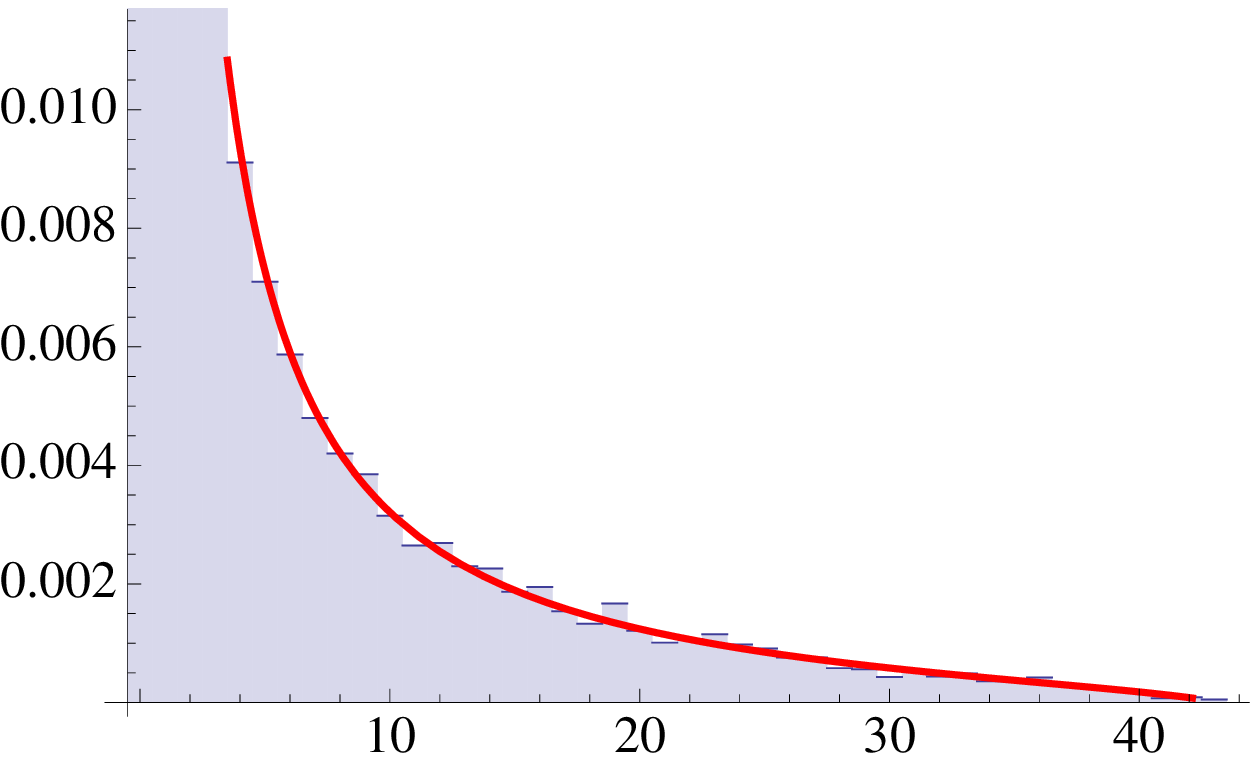}}
\end{picture}
\caption{Spectral density for square matrices for $M=10$, $15$, and $20$ (left to right, up to bottom. Red line plots correspond to the theoretical prediction of formula \eref{eq:sM}). For the numerical diagonalisation,  we have used square matrices of size 1000 and averaged over 1000 samples. In the last plot we have estimated numerically the upper endpoint using square matrices of size 1000 and averaging over a 100 samples. We have also plotted the error bars indicating the typical fluctuations of the largest eivenvalue. As we can see, as $M$ increases the typical fluctuation also increases. This is due to the fact that the fluctuations are amplified when the number of matrix multiplications increases.}
\label{fig:generalM}
\end{center}
\end{figure}

Finally, it is possible to find an exact expression for the spectral density for any $M$ and square matrices by using the method of series inversion. Indeed, after changing variables $u=1+v$, the polynomial equation \eref{eq:polynomial} can be written as follows:
\beeqn{
t=\frac{u}{\phi(u)}\,,\quad \phi(u)\equiv (u-1)^{\frac{1}{M+1}}\,,\quad t=z^{\frac{1}{M+1}}e^{\frac{2\pi k i}{M+1}}\,,\quad k=0,\ldots,M\,.
}
The idea is then to obtain $u(t)$ by inverting the series. We do this using  Lagrange-Burm\"ann formula: $u(t)=\sum_{n=1}^\infty\frac{1}{n}[w^{n-1}]\phi^n(w) t^n$, where here the symbol $[z^{n}]f(z)$ is an operator that gives the $n$-th coefficient of the power series of $f(z)$. After some algebra, going back to the variable $\sigma(1)$, choosing the $k$ that gives the spectral density (it turns out to be $k=0$) we end up with
\beeqn{
\hspace{-2cm}\rho_{M}(\lambda)=\frac{1}{\pi}\left\{\sum_{n=1}^\infty\frac{(-1)^{n-1}}{n}\left(\begin{array}{c}\frac{n}{M+1}\\n-1\end{array}\right) \lambda^{\frac{n}{M+1}-1}\sin\left(\frac{\pi n}{M+1}\right)\right\}\bm{1}_{\lambda\in\left[0, M\left(1+\frac{1}{M}\right)^{M+1}\right]}\,.
\label{eq:sM}
}
This result, which is also called Fuss-Catalan distribution, was also derived in \cite{Penson2011}. Let us check that indeed we recover the previous cases:
\begin{itemize}
\item For $M=1$ we have
\beeq{
\rho_1(w)=\frac{1}{2\pi}\sqrt{\frac{4-w}{w}}\,.
}
\item For $M=2$ we have
\beeqn{
\hspace{-2cm}\rho_2(w)=\frac{1}{2\sqrt{3}\pi w}\left(w^{2/3}\,_{2}F_{1}\left[\frac{1}{6},\frac{2}{3};\frac{4}{3};\frac{4w}{27}\right)-3w^{1/3}\,_{2}F_{1}\left(-\frac{1}{6},\frac{2}{3};\frac{2}{3};\frac{4w}{27}\right)\right]
}
which, after expressing the hypergeometric functions using basic ones, we obtain the results obtained previously.
\item For $M=3$ we obtain a rather involved expression in terms of generalised hypergeometric functions: 
\beeqn{
\hspace{-2cm}\rho_3(w)&=\frac{1}{65536 \pi w\sqrt{2}} \Bigg(65536 \sqrt[4]{w} \, _6F_5\left(-\frac{1}{24},\frac{1}{8},\frac{7}{24},\frac{11}{24},\frac{5}{8},\frac{19}{24};\frac{1}{4},\frac{3}{8},\frac{1}{2},\frac{3}{4},\frac{7}{8};\frac{729 w^2}{65536}\right)\\
\hspace{-2cm}&-16384 \sqrt{2} \sqrt{w} \, _6F_5\left(\frac{1}{12},\frac{1}{4},\frac{5}{12},\frac{7}{12},\frac{3}{4},\frac{11}{12};\frac{3}{8},\frac{1}{2},\frac{5}{8},\frac{7}{8},\frac{9}{8};\frac{729 w^2}{65536}\right)\\
\hspace{-2cm}&-128 \sqrt{2}w^{3/2} \, _6F_5\left(\frac{7}{12},\frac{3}{4},\frac{11}{12},\frac{13}{12},\frac{5}{4},\frac{17}{12};\frac{7}{8},\frac{9}{8},\frac{11}{8},\frac{3}{2},\frac{13}{8};\frac{729 w^2}{65536}\right)\\
\hspace{-2cm}&-2048 w^{3/4} \, _6F_5\left(\frac{5}{24},\frac{3}{8},\frac{13}{24},\frac{17}{24},\frac{7}{8},\frac{25}{24};\frac{1}{2},\frac{5}{8},\frac{3}{4},\frac{9}{8},\frac{5}{4};\frac{729 w^2}{65536}\right)\\
\hspace{-2cm}&-224 w^{5/4} \, _6F_5\left(\frac{11}{24},\frac{5}{8},\frac{19}{24},\frac{23}{24},\frac{9}{8},\frac{31}{24};\frac{3}{4},\frac{7}{8},\frac{5}{4},\frac{11}{8},\frac{3}{2};\frac{729 w^2}{65536}\right)\\
\hspace{-2cm}&-39 w^{7/4} \, _6F_5\left(\frac{17}{24},\frac{7}{8},\frac{25}{24},\frac{29}{24},\frac{11}{8},\frac{37}{24};\frac{9}{8},\frac{5}{4},\frac{3}{2},\frac{13}{8},\frac{7}{4};\frac{729 w^2}{65536}\right)\Bigg)\,.
}
It can be checked that this again agrees with the expression found in the previous section.
\end{itemize}
 It is also worth noticing that given a generalized Hypergeometric function $\,_pF_q(a_1,\ldots,a_p;b_1,\ldots,b_q;z)$ with $p=q+1$, the series  converges for $|z|\leq1$. This completely agrees with the formula \eref{eq:bsc} of the upper endpoint we found previously: for $M=2$ we have that $|4w/27|$ or $|w|\leq 27/4$, while for $M=2$ $|729w^2/65536|$ yielding $|w|\leq 256/27$.\\
In figure \ref{fig:generalM}, we have plotted our theoretical findings for $M=10 $, $15$ and $20$ for square matrices and compared them with results from numerical diagonalisation. Moreover, we have numerically checked the formula \eref{eq:bsc} for the upper endpoint.

\section{Summary, part II and other future work}
\label{confw}
In this work we have tackled the problem of the spectral density of products of Wishart diluted random matrices using the cavity method. We have focused in this part in analysing the dense limit. This has allowed us to make contact with already known results and to bring to light new ones.\\
In part II we will show that a polynomial equation similar to eq. \eref{eq:polynomial} can also be derived for random regular graphs. This  allows us to obtain exact results for  $M$ up to three and also permits some humble exact resuls for general values of $M$. For general diluted graphs, the spectral density can only be obtained by solving the self-consistency equations numerically by using either population dynamics or belief-propagation.\\
Apart of the content of part II, there are other research lines that we are currently persuing. As an instance, we are exploring in which way correlation between matrices impacts the statistical properties derived from this type of ensembles \cite{Perez2014}, and extensions of the present work to products of non-Hermitian diluted matrices \cite{Perez2014b}.

\ack
IPC would like to thank Jesper Ipsen for correcting our second numerical observation and pointing out to relevant literature, making this a better rounded paper.

\newpage
\bibliographystyle{iopart-num}
\bibliography{bib}

\newpage
\appendix
\section{Derivation of polynomial equation}
\label{ap:polynomial}
In the cavity equations we introduce the new variables $u(t)=\sigma(t)/(J^2(t-1)\sigma(t-1))$ and $v(t)=\sigma(t)\Sigma(t)$. Upon also defining $\sigma(0)=1/(z J^2(0))$ we obtain the following eqautions:
\beeqn{
v(t)&=\frac{\alpha_{t}}{\alpha_{t-1}}\left(1-u(t+1)\right){ with } t=1,\ldots,M-1\\
v(t)+u(t)&=1{ with } t=1,\ldots,M\\
\Sigma(M)&=-\frac{\alpha_{M}}{\alpha_{M-1}}\frac{J^2(M)}{1- J^2(M)\sigma(M)}
}
Using these new variables the derivation of the polynomial equation \eref{eq:polynomial} becomes fairly simple: $u(t)$ gives a recursion for $\sigma(t)$, whose solution is at $t=M$:
\beeq{
\sigma(M)=\sigma(0)\prod_{s=1}^{M}J^2(s-1) u(s)
\label{eq:recur}
}
The solutions for $v(t)$ and $u(t)$ obviously are
\beeqn{
v(t)=\frac{v(1)}{\alpha_{t-1}}\,,\quad u(t)=1-\frac{1-u(1)}{\alpha_{t-1}}
}
Moreover for $\sigma(M)$ we find that
\beeqn{
\sigma(M)=\frac{1-u(1)}{J^2(M)\alpha_M\left[\frac{1-u(1)}{\alpha_{M}}-1\right]}
}
Using all these results back into eq. \eref{eq:recur} we write
\beeqn{
\frac{1-u(1)}{J^2(M)\alpha_M\left[\frac{1-u(1)}{\alpha_{M}}-1\right]}=\sigma(0)\prod_{s=1}^{M}J^2(s-1) u(s)
}
After arranging terms and recalling that $u(1)=z\sigma(1)$ and $\sigma(0)=1/(z J^2(0)$ we finally obtain
\beeqn{
\mathcal{J}^2_M\prod_{s=1}^{M+1}\left(1+\frac{\alpha_M}{\alpha_{s-1}}v\right)=vz\,,\quad v\equiv -\frac{1-z\sigma(1)}{\alpha_{M}}\,,\quad\mathcal{J}_M\equiv\prod_{s=1}^{M}J(s)
}

\section{Case $M=2$}
\label{ap:casem2}
In this case we have the following cubic polynomial equation  $a\sigma^3(1)+b\sigma^2(1)+c\sigma(1)+d=0$ with coefficients:
\beeq{
\hspace{-2.5cm}d&=\frac{z}{\alpha_2}\,,\quad c=\frac{z(-\alpha_1 z+(1-\alpha_1)(1-\alpha_2))}{\alpha_1\alpha_2}\,,\quad b=\frac{z^2(\alpha_1+\alpha_2-2)}{\alpha_1\alpha_2}\,,\quad a=\frac{z^3}{\alpha_1\alpha_2}\,.
}
There are various ways to write down the solution to a cubic equation. Here we choose the standard one and write the three solutions as $x_k = - \frac{1}{3a}\left(b + u_k C + \frac{\Delta_0}{u_kC}\right)$ for $k=1$, $2$ and $3$, and where $u_1 = 1$, $u_2 = \frac{-1 + i\sqrt{3}}{2}$ and $u_3 = \frac{-1 - i\sqrt{3}}{ 2}$ and where we have defined:
\beeq{
\hspace{-2.5cm}C = \sqrt[3]{\frac{\Delta_1 + \sqrt{\Delta_1^2 - 4 \Delta_0^3}}{2}}\,,\quad \Delta_0 = b^2-3 a c\,,\quad  \Delta_1 = 2 b^3-9 a b c+27 a^2 d\,.
}
After some tedious algebra one can cancel some factors yielding
\beeqn{
x_k &= - \frac{1}{3 z}\left(\gamma-3 + u_k\tilde{C}  + \frac{\tilde{\Delta}_0}{u_k\tilde{C}}\right)
}
with 
\beeqn{
\tilde{C}&=\sqrt[3]{\frac{\tilde{\Delta}_1 + \sqrt{\tilde{\Delta}_1^2 - 4 \tilde{\Delta}_0^3}}{2}}\,,\quad \tilde{\Delta}_0=3 z\alpha_1+\gamma^2+3(1-\gamma_1\gamma_2)\\
\tilde{\Delta}_1&=(2\gamma^3-9(\alpha_1\gamma_1+\alpha_2\gamma_2+\alpha_1\alpha_2(\gamma-1))+9 z\alpha_1\gamma)\,,
}
where we have defined $\gamma=1+\alpha_1+\alpha_2$ and $\gamma_i=1+\alpha_i$. It is important to notice that factor $1/z$ in the expression of $x_k$ allows us to use distribution theory comfortably to get the continuous part of the spectral density as well as the weight of Dirac delta. Now, a numerical check reveals that the spectral density is given by $k=3$, while the endpoints are given by the roots of $\tilde{\Delta}_1^2 - 4 \tilde{\Delta}_0^3= 0$. In turns out that the latter is a cubic polynomial equation  $az^3 +bz^2+cz +d=0$ with coefficients
\beeqn{
\hspace{-2cm}d&=-27 (-1 + \alpha_1)^2 (\alpha_1 - \alpha_2)^2 (-1 + \alpha_2)^2\,,\\
\hspace{-2cm}c&=-216 \alpha_1 (1 + \alpha_1 + \alpha_2)^2 (1 - (1 + \alpha_1) (1 + \alpha_2)) - 324 \alpha_1 (1 - (1 + \alpha_1) (1 + \alpha_2))^2\\
\hspace{-2cm}&- 162 \alpha_1 (1 + \alpha_1 + \alpha_2) (\alpha_1 (1 + \alpha_1) + \alpha_2 (1 + \alpha_2) + \alpha_1 \alpha_2 (\alpha_1 + \alpha_2))\,,\\
\hspace{-2cm}b&=-27 \alpha_1^2 (1 + \alpha_1 + \alpha_2)^2 - 324 \alpha_1^2 (1 - (1 + \alpha_1) (1 + \alpha_2))\,,\\
\hspace{-2cm}a&=-108 \alpha_1^3\,.
}

\section{Case $M=3$}
\label{ap:casem3}
In this case we write the quartic polynomial equation as  $a\sigma^4(1)+b\sigma^3(1)+c\sigma^2(1)+d\sigma(1)+e=0$ with coefficients
\beeqn{
\hspace{-1cm}a&=\frac{\lambda^4}{\alpha_1\alpha_2\alpha_3}\,,\quad b=\frac{(-3 + \alpha_1 + \alpha_2 + \alpha_3) \lambda^3}{\alpha_1 \alpha_2 \alpha_3}\,,\quad\\
\hspace{-1cm} c&=\frac{(3 + \alpha_2 (-2 + \alpha_3) - 2 \alpha_3 + \alpha_1 (-2 + \alpha_2 + \alpha_3)) \lambda^2}{\alpha_1 \alpha_2 \alpha_3}\,,\\
\hspace{-1cm}d&=\frac{\lambda ((-1 + \alpha_1) (-1 + \alpha_2) (-1 + \alpha_3) - \alpha_1 \alpha_2 w)}{\alpha_1 \alpha_2 \alpha_3}\,,\quad e=\frac{\lambda}{\alpha_3}\,.
}
Using the standard expression for the four solutions one ends up with the following expression:
\beeqn{
\hspace{-1cm}x_{\sigma_1,\sigma_2}&= -\frac{-4 +\gamma}{4\lambda} - \sigma_1\frac{1}{\lambda}\mathcal{S}(\lambda) +\sigma_2\frac{1}{2\lambda}\sqrt{-4\mathcal{S}^2(\lambda) - \frac{\gamma^2-4\gamma_2}{4} +\sigma_1 \frac{1}{8}\frac{Q_1(\lambda)}{\mathcal{S}(\lambda)}}\\
\hspace{-1cm}\mathcal{S}(\lambda)&=\frac{1}{2}\sqrt{- \frac{\gamma^2-4\gamma_2}{12}+\frac{1}{3}\left(\sqrt[3]{\frac{P_2(\lambda) + \sqrt{P_4(\lambda)}}{2}} + \frac{P_1(\lambda)}{\sqrt[3]{\frac{P_2(\lambda) + \sqrt{P_4(\lambda)}}{2}}}\right)}
}
with $\sigma_1,\sigma_2\in\{-1,1\}$ and $\gamma=1+\alpha_1+\alpha_2+\alpha_3$ $\gamma_2=1+\alpha^2_1+\alpha^2_2+\alpha^2_3$. Here $Q_1(\lambda)$, $P_1(\lambda)$, $P_2(\lambda)$, and $P_4(\lambda)$ are the following polyonomials
\beeqn{
\hspace{-2cm}Q_1(\lambda)&=  \alpha_1^3 + (-1 + \alpha_2 - \alpha_3) (1 + \alpha_2 - \alpha_3) (-1 + \alpha_2 + \alpha_3) -   \alpha_1^2 (1 + \alpha_2 + \alpha_3) \\
\hspace{-2cm}&- \alpha_1 (\alpha_2^2 + (-1 + \alpha_3)^2 - 2 \alpha_2 (1 + \alpha_3 - 4 \lambda))\,,\\
\hspace{-2cm}P_1(\lambda)&=  \alpha_3^2 - \alpha_2 \alpha_3 (1 + \alpha_3) + \alpha_2^2 (1 + (-1 + \alpha_3) \alpha_3) -   \alpha_1 (\alpha_2 (1 + \alpha_2) + \alpha_3 \\
\hspace{-2cm}&+ (-6 + \alpha_2) \alpha_2 \alpha_3 + (1 + \alpha_2) \alpha_3^2)\\
\hspace{-2cm}& +\alpha_1^2 (1 + \alpha_2^2 + (-1 + \alpha_3) \alpha_3 - \alpha_2 (1 + \alpha_3 - 3 \lambda)) + 3 \alpha_1 \alpha_2 (1 + \alpha_2 + \alpha_3) \lambda\,,\\
\hspace{-2cm}P_2(\lambda)& = (\alpha_2 (-2 + \alpha_3) + \alpha_3) (\alpha_2 + (-2 + \alpha_2) \alpha_3) (-\alpha_3 +\alpha_2 (-1 + 2 \alpha_3)) \\
\hspace{-2cm}&-3 \alpha_1^2 (\alpha_3 + (-4 + \alpha_3) \alpha_3^2 + \alpha_2^3 (1 + \alpha_3)\\
\hspace{-2cm}& +2 \alpha_2^2 (-2 + \alpha_3 - 2 \alpha_3^2) + \alpha_2 (1 + \alpha_3) (1 + \alpha_3 + \alpha_3^2)) \\
\hspace{-2cm}&+ 9 \alpha_1^2 \alpha_2 (1 + \alpha_2^2 + (-3 + \alpha_3) \alpha_3 - 3 \alpha_2 (1 + \alpha_3)) \lambda\\
\hspace{-2cm} &+  27 \alpha_1^2 \alpha_2^2 \lambda^2 +\alpha_1^3 (2 \alpha_2^3 + (-2 + \alpha_3) (1 + \alpha_3) (-1 + 2 \alpha_3) -3 \alpha_2 (1 + (-4 + \alpha_3) \alpha_3) \\
\hspace{-2cm}&- 3 \alpha_2^2 (1 + \alpha_3 - 3 \lambda) + 9 \alpha_2 (1 + \alpha_3) \lambda)\\
\hspace{-2cm}& + 3 \alpha_1 (-\alpha_3^2 (1 + \alpha_3) + \alpha_2^3 (-1 + 3 \lambda + \alpha_3 (4 - \alpha_3 + 3 \lambda)) \\
\hspace{-2cm}&+\alpha_2 \alpha_3 (4 + 3 \lambda + \alpha_3 (-2 + 4 \alpha_3 + 3 \lambda))\\
\hspace{-2cm}& -\alpha_2^2 (1 - 3 \lambda + \alpha_3 (2 + \alpha_3 (2 + \alpha_3 - 3 \lambda) + 9 \lambda)))\,,\\
\hspace{-2cm}P_4(\lambda) &= P^2_2(\lambda)-4 P^3_1(\lambda)\,.
}
A numerical check reveals that the spectral density is given by the choice $\{\sigma_1,\sigma_2\}=\{-1,1\}$ leading to the reported expression of the spectral density for $M=3$.

\end{document}